\documentclass[a4paper]{elsarticle}

\usepackage[footnotesize]{caption} 
\usepackage[colorlinks=true]{hyperref}
\usepackage{bookmark} 

\usepackage[usenames]{xcolor}
\usepackage{amsmath}
\usepackage{amssymb}
\usepackage{algorithm}
\usepackage{algpseudocode}
\usepackage{listings}
\usepackage{graphicx}
\usepackage{subfig}

\usepackage{booktabs}
\usepackage{multirow}
\usepackage{dcolumn}

\begin{document}

\begin{frontmatter}

\title{PetIGA: A Framework for High-Performance Isogeometric Analysis}

\author[KAUST,CONICET]{L.~Dalcin\corref{corauthor}}
\ead{dalcinl@gmail.com}
\author[ORNL]{N.~Collier}
\ead{nathaniel.collier@gmail.com}
\author[KAUST1]{P.~Vignal}
\ead{philippe.vignal@kaust.edu.sa}
\author[KAUST]{A.M.A.~C\^ortes}
\ead{adrimacortes@gmail.com}
\author[KAUST]{V.M.~Calo}
\ead{vmcalo@gmail.com}

\address[KAUST]{%
Center for Numerical Porous Media (NumPor)\\
Applied Mathematics \& Computational Science and Earth Science \& Engineering\\
King Abdullah University of Science and Technology (KAUST)\\
Thuwal, Saudi Arabia}
\address[CONICET]{%
Centro de Investigaci\'on de M\'etodos Computacionales (CIMEC)\\
Consejo Nacional de Investigaciones Cient\'{\i}ficas y T\'ecnicas (CONICET)\\
Universidad Nacional del Litoral (UNL)\\
Santa Fe, Argentina}
\address[ORNL]{%
Computer Science and Mathematics Division\\
Oak Ridge National Laboratory\\
Oak Ridge, TN, USA}
\address[KAUST1]{%
Center for Numerical Porous Media (NumPor)\\
Materials Science \& Engineering\\
King Abdullah University of Science and Technology (KAUST)\\
Thuwal, Saudi Arabia}
\cortext[corauthor]{Corresponding author}

\begin{abstract}
We present PetIGA, a code framework to approximate the solution of
partial differential equations using isogeometric analysis. PetIGA can
be used to assemble matrices and vectors which come from a Galerkin
weak form, discretized with Non-Uniform Rational B-spline basis
functions. We base our framework on PETSc, a high-performance library
for the scalable solution of partial differential equations, which
simplifies the development of large-scale scientific codes, provides
a rich environment for prototyping, and separates parallelism from
algorithm choice. We describe the implementation of PetIGA, and
exemplify its use by solving a model nonlinear problem. To illustrate
the robustness and flexibility of PetIGA, we solve some challenging
nonlinear partial differential equations that include problems in both
solid and fluid mechanics. We show strong scaling results on
up to $4096$ cores, which confirm the suitability of PetIGA for large
scale simulations.
\end{abstract}


\end{frontmatter}

\clearpage
\section{Introduction}\label{sec:introduction}
Isogeometric analysis, a finite element method originally proposed in
2005~\cite{Hughes2005,Cottrell2009}, was originally motivated by the
desire to find a technique for solving partial differential equations
which would simplify or remove the problem of converting geometric
descriptions for discretizations in the engineering design process.
Once a design is born inside a Computer Aided Design (CAD) program,
converting the CAD representation to an analysis-suitable form usually
is a bottleneck of the engineering analysis process. Isogeometric
methods aim to use CAD representations directly by using the
Non-Uniform Rational B-spline~(NURBS) basis, circumventing
the need to generate an intermediate geometrical description. The term
\textit{isogeometric} reflects that as the finite element space is
refined, the geometrical representation can be preserved
exactly. NURBS technologies have been used in CAD for decades due to
their properties, particularly the smoothness and ability to represent
conic sections. The key insight of isogeometric analysis is to use the
geometrical map of the NURBS representation as a basis for the push
forward used in analysis. This allows isogeometric modeling to advance
where predecessors have found
limitations~\cite{Bercovier1987,Sheffer1998,Sheffer1999,Hollig2003}.

In addition to the geometrical benefits, the basis is also well-suited
to solving higher-order partial differential equations, such as the
ones related to phase-field
problems~\cite{Gomez2008,Gomez2010,Vignal2015b} or large deformation
shell formulations~\cite{Benson2011,Kiendl2015,Bouclier2015}.
Classical finite element spaces use basis functions which are $C^0$
continuous across element boundaries, making them unsuitable for
higher-order problems using a primal Galerkin formulation. The
NURBS-based spaces may be constructed to possess arbitrary degrees of
inter-element continuity for any spatial dimension. These higher-order
continuous basis functions have been
numerically~\cite{Akkerman2008,Akkerman2011,Buffa2011,Evans2013} and
theoretically~\cite{Veiga2011,Evans2009} observed to possess superior
approximability per degree of freedom when compared to their $C^0$
counterparts. However, when used to discretize a Galerkin weak form,
the higher-order continuous basis functions have also been shown to
result in linear systems which are more expensive to solve with
multifrontal direct solvers~\cite{Collier2014,Collier2012} and
iterative solvers~\cite{Collier2013}. These results motivate the
development of efficient, scalable software frameworks which can
mitigate the increase of cost.

This paper describes a scalable implementation of tensor-product,
NURBS-based isogeometric analysis. Despite the fact that writing every
piece of code from scratch is still common practice in research
groups, we believe that reusable software should become the norm in
the scientific community. Otherwise, years of accumulated expertise in
the field are disregarded~\cite{Baxter2012,Wilson2008}, which is
inconsistent with the open and incremental nature of scientific
discovery. However, the choice to depend on existent software should
not be made lightly. Software libraries require maintenance and
development to adapt to changing software and hardware
requirements. This means that they require both personnel and funding
to continue to exist. Furthermore, interfacing with other libraries
can require development work on the library side. Without willing
developers to support and assist third parties in using their
libraries, the process can become cumbersome. We believe the benefits
outweigh the risks, yet these are factors to consider when one plans
to reuse the work of others.

PETSc~\cite{petsc,petsc-manual,petsc-efficient}, the \emph{Portable
Extensible Toolkit for Scientific Computation}, is a collection of
algorithms and data structures for the solution of scientific
problems, particularly those modeled by partial differential
equations. PETSc is applicable to a wide range of problem sizes,
including extreme large-scale simulations, where high-performance
parallel computation is a must. PETSc uses the message-passing
interface (MPI) model for communication, but provides high-level
interfaces with collective semantics so that typical users rarely
have to make message-passing calls directly.

PETSc provides a rich environment for modeling scientific problems as
well as for rapid algorithm design and prototyping. The library
enables easy customization and extension of both algorithms and
implementations. This approach promotes code reuse and
flexibility. PETSc is object-oriented in style, with components that
may be changed via a command-line interface at runtime. These
components include:
\begin{itemize}
\item Index sets to describe permutations, indexing, renumbering, and communication patterns;
\item Matrices and vectors that provide basic linear algebra abstractions;
\item Krylov subspace methods and preconditioners that include multigrid and sparse direct solvers;
\item Nonlinear solvers and time stepping algorithms; and
\item Distributed arrays for parallelizing structured grid-based problems.
\end{itemize}

PETSc is also designed to be highly modular, enabling the
interoperability with specialized parallel libraries like
{Hypre}~\cite{hypre}, {Trilinos/ML}~\cite{trilinos},
{MUMPS}~\cite{mumps}, and others through a unified interface. Other
scientific packages geared towards solving partial differential
equations use components from PETSc (for example
{deal.II}~\cite{dealII}, {FEniCS}~\cite{fenics},
{libMesh}~\cite{libmesh}, and {PETSc-FEM}~\cite{petscfem}).

In this paper we describe an approach to reuse PETSc algorithms and
data structures to obtain a high-performance framework designed for
isogeometric analysis. We implement parallel vector and matrix
assembly using PETSc data structures and interface into PETSc's wide
range of solvers. In section~\ref{sec:implementation}, we detail the
implementation as well as the features of the framework. In
section~\ref{sec:example}, we tackle a model problem for nonlinear
applications, and go through all the steps to solve it using our
framework. Finally we showcase applications in
section~\ref{sec:applications} and discuss performance results in
section \ref{sec:performance}. We call our framework PetIGA. It is
freely available~\cite{petiga} and under active development.


\section{Implementation}\label{sec:implementation}
\subsection{B-spline basis functions}

Let $\Xi=\{\xi_0,\dots,\xi_m\}$ be a non-decreasing sequence of real
numbers, i.e., $\xi_i \leq \xi_{i+1}, i=0,\dots,m-1$. The $\xi_i$ are
called \emph{knots} and $\Xi$ is the \emph{knot vector}. By using the
Cox--de~Boor recursion formula~\cite{Cox1972,DeBoor1972}, the $i$th
B-spline basis function of $p$-degree, $i=0,\dots,m-p-1$, denoted
$B_{i,p}(\xi)$, is defined as
\begin{align}
  B_{i,0}\left(\xi\right) &=
  \begin{cases}
    1\ \ \text{if } \xi_i \le \xi < \xi_{i+1},\\
    0\ \ \text{otherwise,}
  \end{cases}\\
  B_{i,p}\left(\xi\right) &=
  \frac{\xi-\xi_i}{\xi_{i+p}-\xi_i}B_{i,p-1}(\xi)+
  \frac{\xi_{i+p+1}-\xi}{\xi_{i+p+1}-\xi_{i+1}}B_{i+1,p-1}(\xi).
\end{align}
Writing $B_{i,p}$ instead of $B_{i,p}(\xi)$ for brevity, the first
derivative of a basis function is given by
\begin{equation}
  B'_{i,p}(\xi) = \frac{p}{\xi_{i+p}-\xi_i}B_{i,p-1}+\frac{p}{\xi_{i+p+1}-\xi_{i+1}}B_{i+1,p-1}.
\end{equation}
Repeated differentiation of the previous expression produces the
general formula for the $k$th derivative $B^{(k)}_{i,p}(\xi)$ of
$B_{i,p}(\xi)$
\begin{equation}
  B^{(k)}_{i,p}(\xi) = p\left(\frac{B^{(k-1)}_{i,p-1}}{\xi_{i+p}-\xi_i}+\frac{B^{(k-1)}_{i+1,p-1}}{\xi_{i+p+1}-\xi_{i+1}}\right).
\end{equation}
Even though this is the standard way of expressing the basis
functions, we compute basis values and derivatives using the more
computationally efficient algorithms detailed in~\cite{Piegl1995}.

\subsection{Tensor product basis functions}

By using a tensor product structure, a one-dimensional basis can be
extended to multi-dimensions. Here we write a three-dimensional
extension of the one-dimensional B-spline basis. Let %
\begin{equation*}
\Xi        = \{\xi_0,\dots,\xi_{n+p+1}\}, \ %
\mathrm{H} = \{\eta_0,\dots,\eta_{m+q+1}\}, \ %
\mathrm{Z} = \{\zeta_0,\dots,\zeta_{o+r+1}\}
\end{equation*}
be knots vectors which define three sets of one-dimensional B-spline
basis functions %
\begin{equation*}
\{B_{i,p}(\xi),   i=0,\dots,n\}, \ %
\{B_{j,q}(\eta),  j=0,\dots,m\}, \ %
\{B_{k,r}(\zeta), k=0,\dots,o\}
\end{equation*}
of degree $p$, $q$, and $r$, respectively. %
The three-dimensional B-spline basis functions are defined as
\begin{equation}
  M_{ijk}(\xi,\eta,\zeta) = B_{i,p}(\xi)B_{j,q}(\eta)B_{k,r}(\zeta).
\end{equation}
For the sake of notational convenience and dimension independence, we
denote multi-dimensional B-spline basis functions as
$M_A(\boldsymbol{\xi})$ in the following. For the particular case of
three dimensions, $\boldsymbol{\xi}$ denotes the parametric coordinate
$(\xi,\eta,\zeta)$ and $A$ is the global basis index, i.e.,
$A=i+j(n+1)+k(n+1)(m+1)$.

\subsection{NURBS basis functions and derivatives}

Given the B-spline basis functions $M_A(\boldsymbol{\xi})$, we can
define the corresponding NURBS~\cite{Piegl1995} basis as
\begin{equation}\label{eq:rational}
  N_A(\boldsymbol{\xi}) = \dfrac{w_A M_A(\boldsymbol{\xi})}{w(\boldsymbol{\xi})}
\end{equation}
where $w_A$ are the projective weights and the weighting function
appearing in the denominator is
\begin{equation}
  w(\boldsymbol{\xi})=\sum_B w_BM_B({\boldsymbol\xi})
\end{equation}
Omitting the explicit dependence on ${\boldsymbol\xi}$ for notational
convenience, derivatives of $w$ with respect to the parametric
coordinates are simply
\begin{align}
  w_{,\alpha} &= \dfrac{\partial w}{\partial\xi_\alpha} = \sum_B w_B M_{B,\alpha}\\
  w_{,\alpha\beta} &= \dfrac{\partial^2 w}{\partial\xi_\alpha\partial\xi_\beta} = \sum_B w_B M_{B,\alpha\beta}\\
  w_{,\alpha\beta\gamma} &= \dfrac{\partial^3 w}{\partial\xi_\alpha\partial\xi_\beta\partial\xi_{\gamma}} = \sum_B w_B M_{B,\alpha\beta\gamma}
\end{align}
Using the chain rule, the first, second, and third derivatives with
respect to $\xi_\alpha$ of the rational basis function $N_A$ defined
in~\eqref{eq:rational} can be expressed as
\begin{align}
  N_{A,\alpha}&=\dfrac{w_AM_{A,\alpha}-N_Aw_{,\alpha}}{w}\label{eq:drational}\\
  N_{A,\alpha\beta}&=\dfrac{w_AM_{A,\alpha\beta}-N_Aw_{,\alpha\beta}}{w}\notag\\
  &-\dfrac{N_{A,\beta}w_{,\alpha}+N_{A,\alpha}w_{,\beta}}{w}\label{eq:ddrational}\\
  N_{A,\alpha\beta\gamma}&=\dfrac{w_AM_{A,\alpha\beta\gamma}-N_Aw_{,\alpha\beta\gamma}}{w}\notag\\
  &-\dfrac{N_{A,\alpha}w_{,\beta\gamma}+N_{A,\beta}w_{,\alpha\gamma}+N_{A,\gamma}w_{,\alpha\beta}}{w}\notag\\
  &-\dfrac{N_{A,\beta\gamma}w_{,\alpha}+N_{A,\alpha\gamma}w_{,\beta}+N_{A,\alpha\beta}w_{,\gamma}}{w}\label{eq:dddrational}
\end{align}

In~\eqref{eq:rational}, we gave a definition of the rational basis
function in terms of its parametric coordinates. However, when using
the isoparametric concept~\cite{Hughes2000}, we need to express
derivatives in terms of spatial coordinates, not their parametric
counterparts. Higher-order derivatives of the geometrical mapping and
basis functions are not standard in the literature and are required if
one is to solve a higher-order partial differential equations on a
mapped geometry. For the sake of completeness, the derivation of these
derivatives can be found in appendix~\ref{app:ho-sp-ders}.

\subsection{Periodic boundary conditions}

Due to the prevalent use of open knot vectors by the IGA community,
applications which require the enforcement of periodic boundary
conditions typically do so by building a system of constraints on the
coefficients. For example, in~\cite{Liu2013}, the authors detail
constraint equations which enforce $C^1$ periodicity. However, we
prefer to build periodicity into the function space due to its
simplicity and generality. We do this by \emph{unclamping} the knot
vector as in the parlance of~\cite{Piegl1995}. Let
$\Xi=\{\xi_0,\ldots,\xi_m\}$ be an open (or \emph{clamped}) knot
vector which encodes a set of $p$-degree B-spline basis functions
$\{B_{i,p}(\xi),i=0,\dots,n\}$, where $m=n+p+1$. Unclamping the knot
vector for a desired continuity $C^k$ at the boundary,
$0 \leq k \le p-1$, amounts to redefining $k+1$ knot values at the
left and right ends.
\begin{alignat}{2}
  \xi_{k-i} &= \xi_{p} - \xi_{n+1} + \xi_{n-i}, \quad && 0 \le i \le k,\label{eq:unclamp-left}\\
  \xi_{m-k+i} &= \xi_{n+1} - \xi_{p} + \xi_{p+i+1}, \quad && 0 \le i \le k.\label{eq:unclamp-right}
\end{alignat}

\begin{figure}[p]
\centering
\subfloat[$C^0$ periodicity, $\Xi=\{-0.2,0,0,0,0.2,0.4,0.6,0.8,1,1,1,1.2\}$]          {\includegraphics[width=0.80\textwidth]{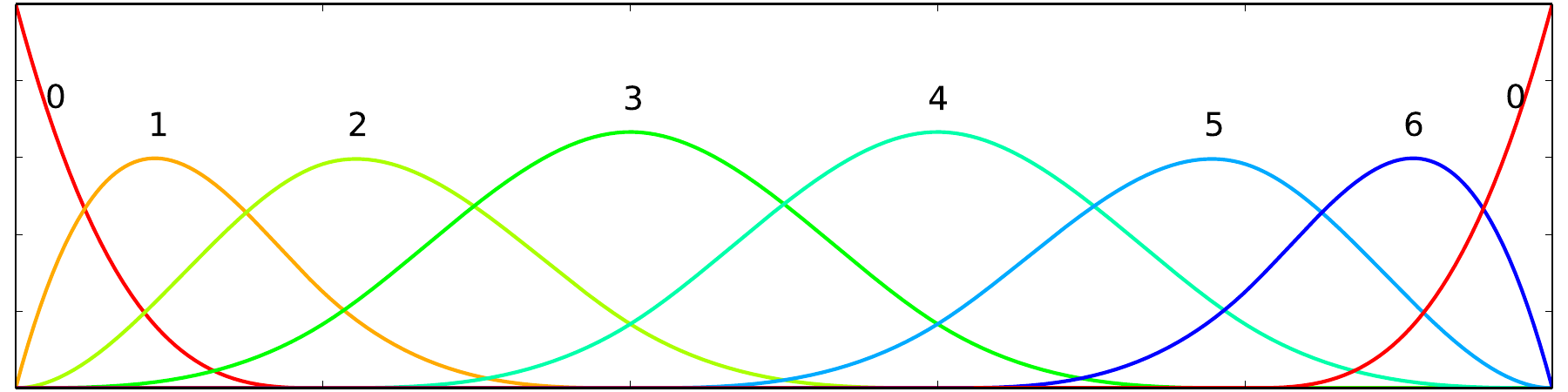}}\\
\subfloat[$C^1$ periodicity, $\Xi=\{-0.4,-0.2,0,0,0.2,0.4,0.6,0.8,1,1,1.2,1.4\}$]     {\includegraphics[width=0.80\textwidth]{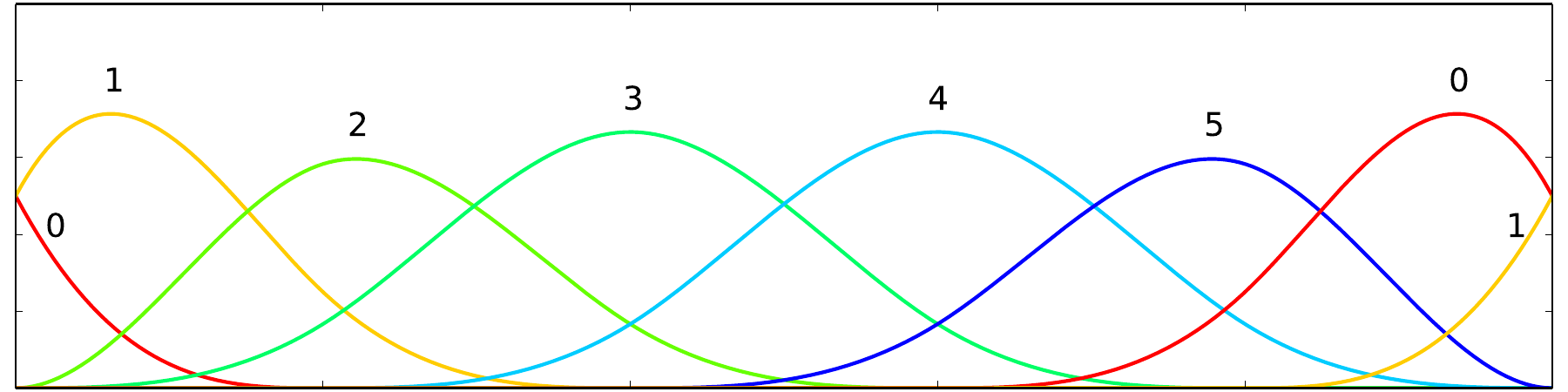}}\\
\subfloat[$C^2$ periodicity, $\Xi=\{-0.6,-0.4,-0.2,0,0.2,0.4,0.6,0.8,1,1.2,1.4,1.6\}$]{\includegraphics[width=0.80\textwidth]{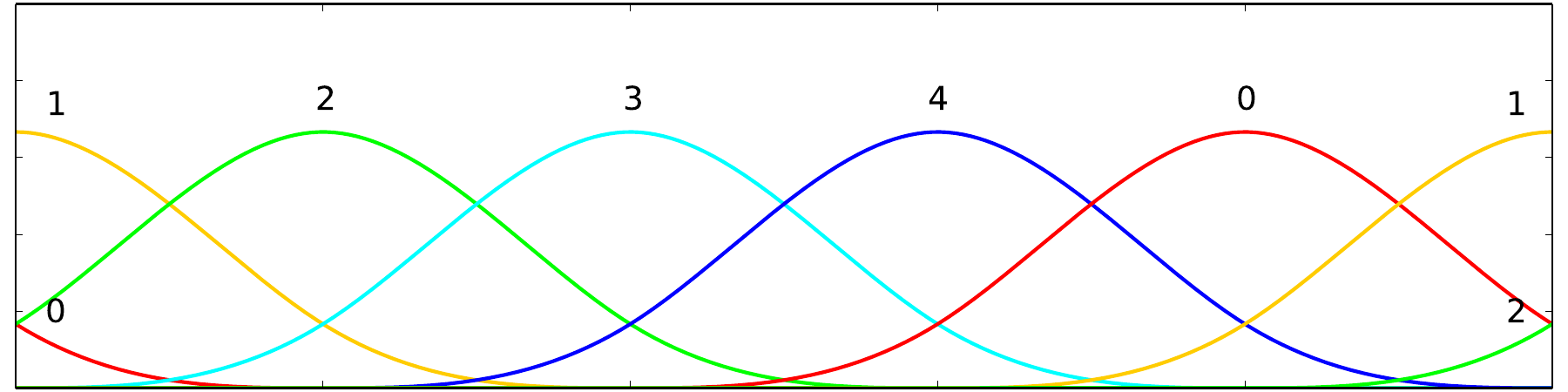}}
\caption{%
  Three cases of periodicity for a $C^2$ cubic B-spline space. In each
  case, the open knot vector was unclamped using
  equations~\eqref{eq:unclamp-left}~and~\eqref{eq:unclamp-right}. Unique
  basis functions are labeled by their global numbering and colored
  distinctly.}\label{fig:unclamp}
\end{figure}

In figure~\ref{fig:unclamp} we present a $C^2$ cubic B-spline space
with varying orders of continuity across the periodic boundary. Each
of these knot vectors was obtained by unclamping the open knot vector
$\Xi=\{0,0,0,0,0.2,0.4,0.6,0.8,1,1,1,1\}$ using
equations~\eqref{eq:unclamp-left}~and~\eqref{eq:unclamp-right}. Each
unique basis is labeled with its global number and colored distinctly
such that basis functions which pass the periodic interface may be
identified.

\begin{figure}[p]
\centering
\includegraphics[width=0.6\textwidth]{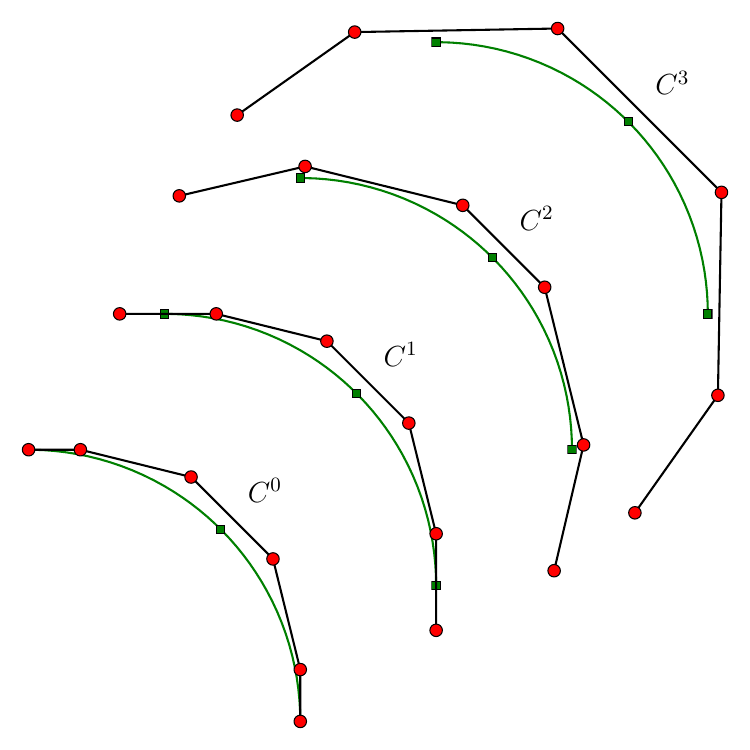}
\caption{%
  A quarter circular arc of degree four repeatedly unclamped by
  algorithm~\ref{alg:unclamp-curve}.}\label{fig:unclamp-curve}
\end{figure}

Equations~\eqref{eq:unclamp-left}~and~\eqref{eq:unclamp-right} are
sufficient when utilizing the basis in the parametric domain. In cases
where the parametric domain is mapped, the original control points of
the mapping which correspond to the open knot vector also need to be
unclamped. In~\cite[p. 577]{Piegl1995}, algorithm A12.1 performs this
operation but is limited to $C^{p-1}$ unclamping. We present a
generalization in algorithm~\ref{alg:unclamp-curve} for $C^k$
unclamping where \texttt{U} is the array of knots and \texttt{Pw} is
the array of control points in homogeneous coordinates. In
figure~\ref{fig:unclamp-curve}, we present the effect of our algorithm
when applied to a quarter circular arc.

\begin{algorithm}[h]
\caption{Pseudocode for unclamping curves.}\label{alg:unclamp-curve}%
\begin{verbatim}
UnclampCurve(n,p,k,U,Pw) {
  /* Input:
      n - index of last basis function
      p - polynomial degree
      k - continuity order
      U, Pw - knot vector and control points
    Output:
      U, Pw (modified in-place) */
  m = n + p + 1; /* index of last knot */
  for (i=0; i<=k; i++) /* Unclamp at left end */
    U[k-i] = U[p] - U[n+1] + U[n-i];
  for (i=p-k-1; i<=p-2; i++)
    for (j=i; j>=0; j--) {
      alpha = (U[p]-U[p+j-i-1])/(U[p+j+1]-U[p+j-i-1]);
      Pw[j] = (Pw[j]-alpha*Pw[j+1])/(1-alpha);
    }
  for (i=0; i<=k; i++) /* Unclamp at right end */
    U[m-k+i] = U[n+1] - U[p] + U[p+i+1];
  for (i=p-k-1; i<=p-2; i++)
    for (j=i; j>=0; j--) {
      alpha = (U[n+1]-U[n-j])/(U[n-j+i+2]-U[n-j]);
      Pw[n-j] = (Pw[n-j]-(1-alpha)*Pw[n-j-1])/alpha;
    }
}
\end{verbatim}
\end{algorithm}

After unclamping a knot vector, imposing periodic boundary conditions
no longer requires the use of constraint equations. Periodicity can
now be embedded into the function space by eliminating redundant basis
functions in one of the domain boundaries. A practical way of dealing
with this approach in a computer code is to properly map indices of
basis functions that were eliminated to the corresponding ones that
remain. In the non-periodic case, the number of basis functions is
$n+1$. Imposing periodic boundary conditions with continuity order $k$
reduces the number of basis functions to $n-k$. Any basis function
index $i$ outside the interval $\left[0,n-k-1\right]$ can be
wrapped-around using the mapping~
$i \mapsto \mathrm{mod}\left(i,n-k\right)$, where
\begin{equation}
  \mathrm{mod}\left(a,b\right) =
  a - b \left\lfloor \frac{a}{b} \right\rfloor\label{eq:map-wrap}
\end{equation}
and the symbol $\lfloor \cdot \rfloor$ denotes the floor function,
i.e., $\lfloor x \rfloor$ is the largest integer not greater than $x$.

\subsection{Adjacency graph}

For numerical methods employing basis functions with local support, it
is important to compute the adjacency graph of interacting degrees of
freedom. This graph contains information required for the proper
preallocation of sparse matrices implemented in compressed sparse row
(CSR) or column (CSC) formats. This preallocation is crucial for
efficient matrix assembly. Furthermore, prior knowledge of the sparse
matrix nonzero pattern enables the use of specialized differentiation
techniques, such as the approximation of Jacobian matrices by colored
finite differences, see section~\ref{sec:num-diff}.

Given an array of knots \texttt{U} encoding a set of B-spline basis
functions of $p$-degree, algorithm~\ref{alg:stencil} computes the
left-most ($\ell$) and right-most ($r$) indices of the basis functions
interacting with the $i$-th basis function. That is, the support of
the basis function $N_{i,p}$ has non-empty intersection with the
support of $N_{j,p}$, for $\ell \leq j \leq r$. In one dimension, all
basis functions with indices in the set $\{\ell,\ell+1,\ldots,r-1,r\}$
are adjacent to the $i$-th basis function. For dimensions higher than
one, the adjacency graph is computed from the index sets
$\{\ell_d,\ell_d+1,\ldots,r_d-1,r_d\}$ for each $d$-dimension using
the tensor-product structure and numbering in lexicographical order.
When imposing periodic boundary conditions with continuity order $k$,
algorithm~\ref{alg:stencil} may return indices outside the interval
$\left[0,n-k-1\right]$. Again, the index mapping given
in~\eqref{eq:map-wrap} has to be used to wrap-around index values.

\begin{algorithm}
\caption{Pseudocode for computing the adjacency graph.}\label{alg:stencil}%
\begin{verbatim}
BasisStencil(i,p,U,l,r) {
  /* Input:
      i - index of basis
      p - polynomial degree
      U - knot vector
     Output: l,r */
      l - index of left-most overlapping basis
      r - index of right-most overlapping basis */
  j = i;
  while (U[j] == U[j+1]) j++;
  l = j - p;
  j = i + p + 1;
  while (U[j] == U[j-1]) j--;
  r = j - 1;
}
\end{verbatim}
\end{algorithm}

\begin{figure}[h]
\centering
\captionsetup[subfigure]{width=0.75\textwidth}
\subfloat[%
A sample representative cubic B-spline basis consisting of four
elements where the inter-element continuity progressively decreases
according to the knot vector $\{0,0,0,0,2,4,4,6,6,6,8,8,8,8\}$.
]{\includegraphics[width=0.8\textwidth]{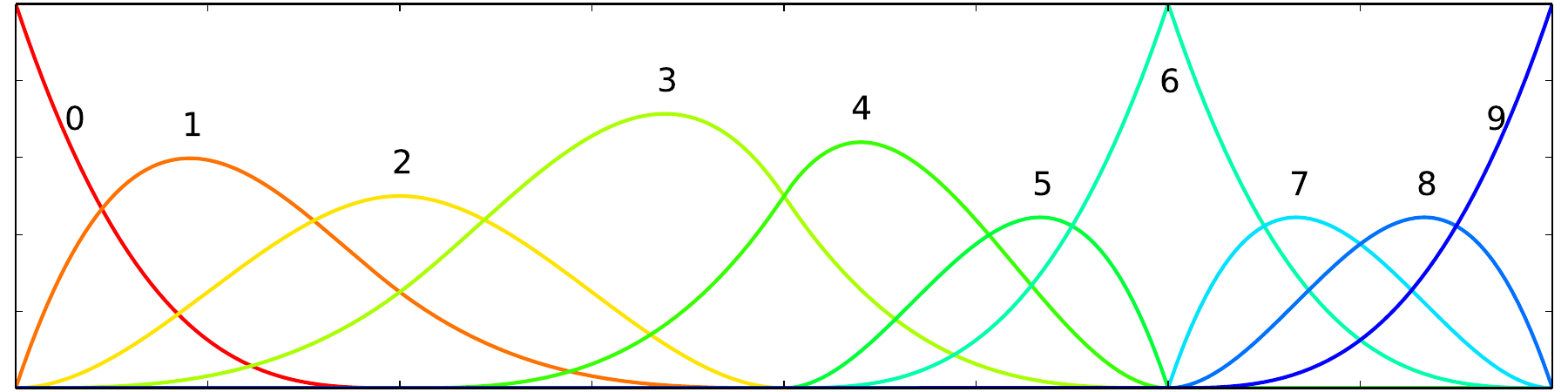}}\\
\smallskip
\subfloat[%
Nonzero structure where the function space represented
above is used as test and trial functions in a Galerkin finite element
method. Nonzero entries are indicated by a square.
]{\makebox[0.75\textwidth%
]{\includegraphics[width=0.50\textwidth]{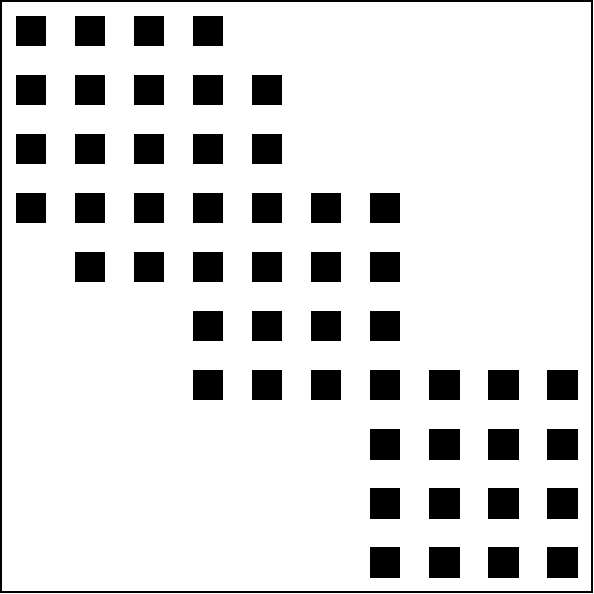}}}
\caption{%
  Generation of the nonzero structure of matrices using
  algorithm~\ref{alg:stencil}.}\label{fig:stencil}
\end{figure}

Figure~\ref{fig:stencil}(a) presents a set of cubic basis functions
which varies in inter-element continuity. Each basis is labeled by a
global number and colored distinctly. Figure~\ref{fig:stencil}(b)
shows the corresponding nonzero pattern for the sparse matrix obtained
by applying algorithm~\ref{alg:stencil} to each basis.

\subsection{Partitioning}

In this section we describe the  parallel partitioning and
inter-process communication PetIGA uses. The ideas and
techniques presented here closely resemble the ones used in PETSc to
handle structured problems (through the \texttt{DMDA} component) as
well as unstructured problems (through the \texttt{DMPlex} component),
see~\cite{petsc-manual}. However, PetIGA implements its own data
structures tailored to the specifics of isogeometric analysis,
particularly the use of higher-continuous, tensor-product polynomial
spaces with possibly arbitrary continuity orders across discrete
domains. In the following, \emph{element} refers to a non-empty knot
span (as defined in~\cite{Cottrell2009}) and \emph{node} refers to an
abstract entity that can encompass control points and weights of a
NURBS geometry, control variables of either a scalar or vector field,
unknown coefficients, and/or residual equation values.

\begin{figure}[p]
\centering
\subfloat[Element grid.
]{\includegraphics[scale=0.80]{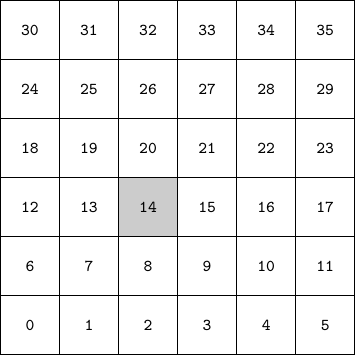}\label{fig:grid-base:elem}}
\\ 
\subfloat[Node grid.
]{\includegraphics[scale=0.80]{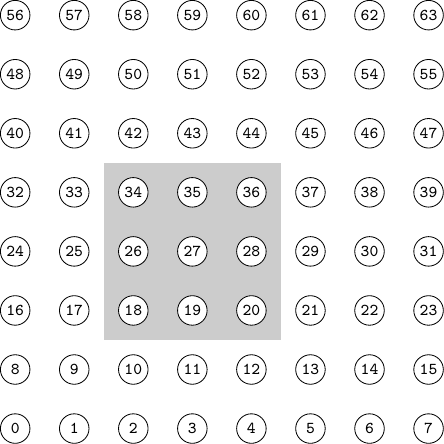}\label{fig:grid-base:node}}
\caption{%
  A two-dimensional quadratic $C^1$ space with $6 \times 6$ elements
  and $8 \times 8$ nodes. The shaded region in
  figure~\ref{fig:grid-base:node} contains the nodes which correspond
  to basis functions with support on the element highlighted in
  figure~\ref{fig:grid-base:elem}.}
\label{fig:grid-base}
\end{figure}

Figure~\ref{fig:grid-base} depicts elements and nodes corresponding to
a two-dimensional quadratic $C^1$ discrete space with $6 \times 6$
elements and $8 \times 8$ nodes.  Both elements and nodes are labeled
with an index which starts at zero and follows a natural numbering. In
figure~\ref{fig:grid-base:node}, a subset of nodes is drawn inside a
shaded region. These nodes correspond to basis functions with support
on the element highlighted in figure~\ref{fig:grid-base:elem}.

\begin{figure}[p]
\centering
\subfloat[Element partition.
]{\includegraphics[scale=0.80]{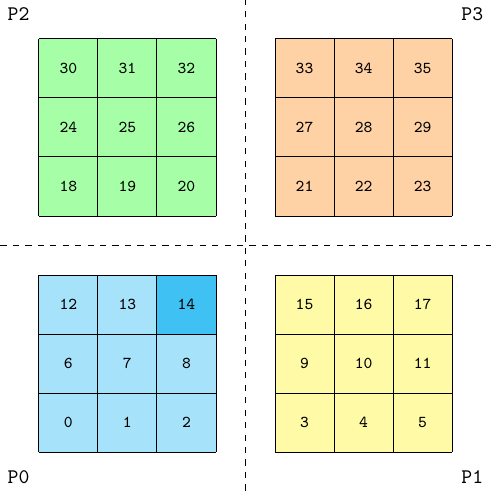}\label{fig:grid-part:elem}}
\\
\subfloat[Node partition and natural numbering.
]{\includegraphics[scale=0.80]{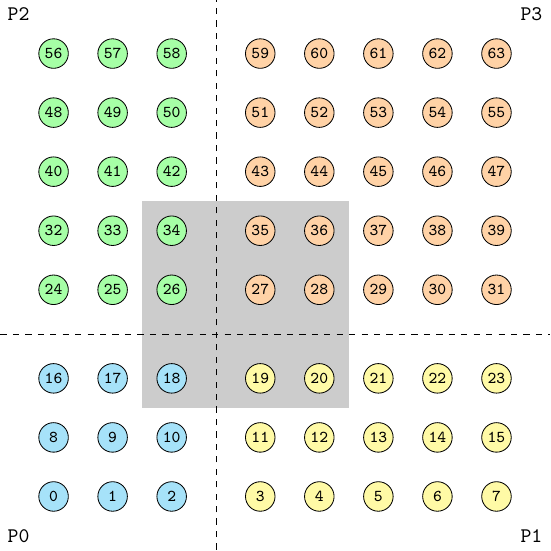}\label{fig:grid-part:node}}
\caption{%
  The element and node grids in figure~\ref{fig:grid-base} are
  partitioned and distributed over a set of four processes
  $\{\texttt{P0,P1,P2,P3}\}$ arranged in a $2 \times 2$ process grid.
  Elements and nodes are labeled according to the natural numbering,
  and assigned a distinct color that depends on which process they
  belong to. The shaded region in figure~\ref{fig:grid-part:node}
  contains the nodes related to the element highlighted in
  figure~\ref{fig:grid-part:elem}.}
\label{fig:grid-part}
\end{figure}

Given a $2 \times 2$ grid of processes labeled from zero to three,
figure~\ref{fig:grid-part} depicts a partition of the sets of elements
and nodes. Figure~\ref{fig:grid-part:elem} highlights a single
element (drawn with a darker color) while
figure~\ref{fig:grid-part:node} highlights its corresponding subset of
nodes (drawn inside a shaded region). The highlighted element is
assigned to process zero whereas the nodes inside the shaded region
are assigned to different processes. Up to this stage, elements and
nodes are still labeled according to the natural numbering defined in
figure~\ref{fig:grid-base}.

\begin{figure}[p]
\centering
\subfloat[Global node grid and global numbering.
]{\includegraphics[scale=0.80]{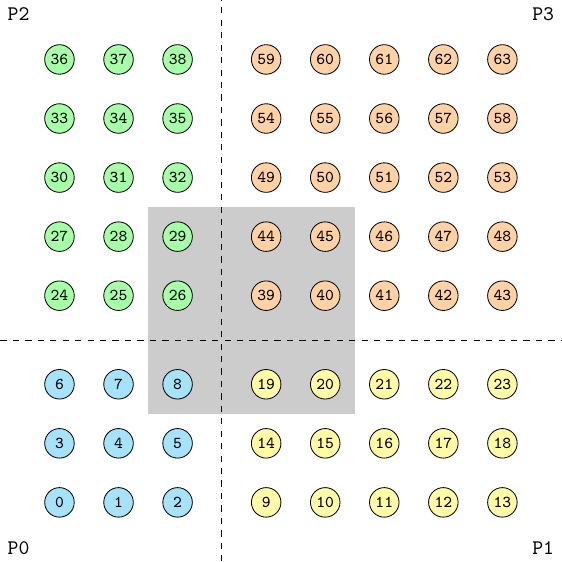}\label{fig:node-global}}
\\
\subfloat[Local node grids and local numberings.
]{\includegraphics[scale=0.80]{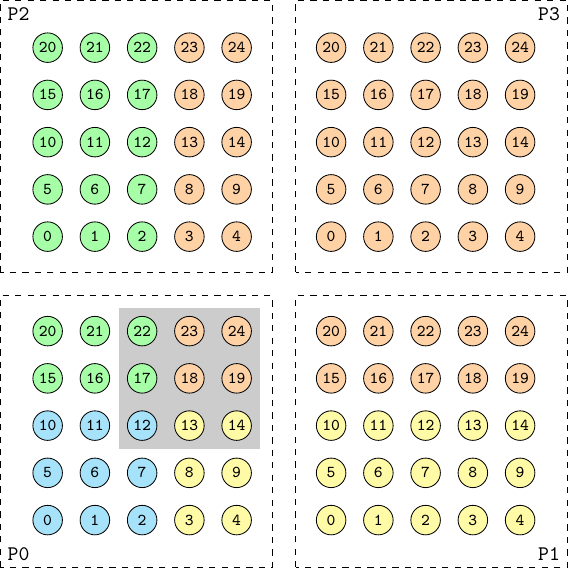}\label{fig:node-local}}
\caption{%
  Global and local node grids. Global nodes are colored according to
  the process they belong to, and labeled following a block-contiguous
  global numbering. Nodes in the local grids are labeled following a
  local numbering. The local grids in processes \texttt{P0},
  \texttt{P1}, and \texttt{P2} contain both strictly-local and ghost
  nodes, while the local grid in process \texttt{P3} only contains
  strictly-local nodes. The shaded region in
  figure~\ref{fig:node-global} is the same as in
  figure~\ref{fig:grid-part:node}. The nodes corresponding to the
  element highlighted in figure~\ref{fig:grid-part:elem} are now fully
  contained within the local grid of process $\texttt{P0}$ as shown in
  figure~\ref{fig:node-local}.}
\label{fig:grid-numbering}
\end{figure}

Figure~\ref{fig:node-global} defines a new node numbering. This global
numbering is block-contiguous and follows the process numbering. The
\emph{natural} numbering in figure~\ref{fig:grid-part:node} and the
\emph{global} numbering in figure~\ref{fig:node-global} define a
bijective $\mathit{natural}~\mapsto~\mathit{global}$ mapping. The
global numbering enhances locality, thus improving overall parallel
performance. However, this numbering depends on the size and layout of
the process grid. As the natural numbering does not depend on the
partitioning, it is better suited for tasks involving data
persistence, such as pre-/post-processing and
checkpoint/restart. PetIGA employs the natural numbering when
datafiles have to be read or written to disk, and uses the
$\mathit{natural}~\mapsto~\mathit{global}$ mapping on the fly to
convert to/from the global numbering.

The subset of nodes drawn inside the shaded region in
figure~\ref{fig:node-global} indicates that computations performed
in the highlighted element in figure~\ref{fig:grid-part:elem} would
require communication with neighboring processes. As most
distributed-memory parallel codes using the message-passing paradigm
and dealing with grid-based methods, PetIGA efficiently handles
inter-process communication by introducing augmented \emph{local
  grids} at each process. Conceptually, a local grid is a sequential
entity that belongs to a single process, while the global grid is a
single entity that is distributed across processes. Within a local
grid, two disjoint subsets of nodes can be distinguished: the
\emph{strictly-local nodes} which replicate the in-process subset of
nodes of the global grid, and the \emph{ghost nodes} which replicate
global nodes assigned to neighboring processes.
Figure~\ref{fig:node-local} depicts the local grids of each process
corresponding to the global grid in figure~\ref{fig:node-global}.
Nodes in the local grids are labeled with an index that starts at zero
and defines the \emph{local} numbering. Strictly-local nodes and ghost
nodes are distinguished by assigning them a color that matches the
process the corresponding global node is assigned to. For example, in
figure~\ref{fig:node-local}, in the local grid of process zero, blue
nodes are strictly local nodes while the rest are ghost nodes. Data
transfer between the global and local grids is managed through
injective $\mathit{local}~\mapsto~\mathit{global}$ mappings defined
within each process. Data associated to strictly-local nodes is
handled with in-process memory transfers, while data associated to
ghost nodes has to be communicated back and forth with neighboring
processes. The communication costs increase with the number of ghost
nodes, which in turn depend on the continuity order at inter-process
interfaces: spaces of continuity $C^k$ lead to a number of ghost nodes
proportional to $k+1$.

\begin{figure}[p]
\centering
\hfill\subfloat[Parallel layout of a global vector distributed across processes.%
]{\includegraphics[scale=1]{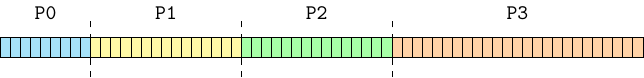}\label{fig:vec-global}}
\\%
\subfloat[Sequential layout of local vectors within each process.%
]{\hspace{2em}\includegraphics[scale=1]{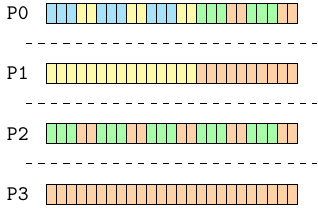}\label{fig:vec-local}}
\\%
\hfill\subfloat[Parallel layout of a global sparse matrix distributed by rows across processes.%
]{\includegraphics[scale=1]{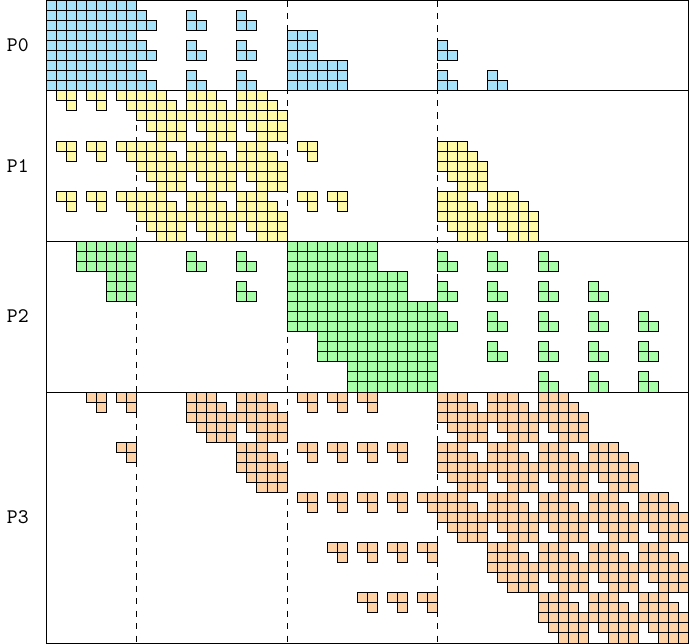}\label{fig:mat-global}}
\caption{%
  Layout of vectors and sparse matrices.}
\label{fig:grid-vecmat}
\end{figure}

Lastly, figure~\ref{fig:grid-vecmat} shows the layout of vectors and
sparse matrices. The global vector in figure~\ref{fig:vec-global} is
the concrete data structure used to store data associated with nodes
from the global grid of figure~\ref{fig:node-global}.  Global vectors
are distributed data structures built out of local one-dimensional
arrays of floating point values. Similarly, the local vectors in
figure~\ref{fig:vec-local} are the concrete counterparts of the local
grids presented in figure~\ref{fig:node-local}. These are nonetheless
sequential data structures which are independent of each other.
Global sparse matrices are distributed by rows across processes, as
shown in figure~\ref{fig:mat-global}. Thanks to the locality induced
by the block-contiguous global numbering, most of the entries within a
process belong to the denser diagonal blocks. This in turn improves
overall performance of global matrix-vector products.

\subsection{Assembly}\label{sec:assembly}

Through the pseudocode in algorithm~\ref{alg:residual}, we describe how
a residual vector coming from the discretization of a nonlinear
partial differential equation is assembled in parallel. We use arrow
symbols to reflect movement of memory, either within or across
processes, and the equality symbol to signify when computation
occurs. Given a global vector $\mathbf{U}$ which constitutes the
current state of the solution to the nonlinear problem, we obtain the
\emph{local vector} $\mathbf{U}_\ell$ by gathering off-process values
through inter-process communication and packing them together with
process-owned values. The steps which follow are standard practice in
finite element codes. We loop over all elements $e$ within a subpatch
and gather solution and geometry coefficients $\mathbf{U}_e$ and
$\mathbf{G}_e$, respectively. Then, we loop over quadrature points
$q$ within the element and compute the set of nonzero basis functions
and their spatial derivatives (which we collectively denote as
$\left\{N_q\right\}$) as well as the determinant of the Jacobian the
geometrical mapping $J_q$. The main difference between applications
lies in the evaluation of the integrand at a quadrature point. This
problem-specific evaluation routine has to be provided by the user. We
then accumulate the quadrature point contributions $\mathbf{F}_q$ into
the element residual vector $\mathbf{F}_e$, taking into account the
quadrature weight $w_q$ and Jacobian determinant $J_q$. Each element
residual $\mathbf{F}_e$ is then assembled into the global residual
vector $\mathbf{F}$. In this step, PETSc automatically handles
off-processor contributions by storing them in a cache which we
designate as $\mathbf{F}_{\text{cache}}$. Finally, after the
completion of the element loop, the cached contributions are
communicated and assembled into the global vector $\mathbf{F}$.
Matrices are assembled in a similar fashion. Again, the user only has
to provide the problem-specific evaluation routine computing the
integrand.

This approach to abstraction hides from the user the details of sparse
matrix and vector assembly as well as parallelism. In turn,
application codes are shorter and easier to read as they only contain
code relevant to the physics of the modeled problem.

\renewcommand{\algorithmicforall}{\textbf{for each}}
\begin{algorithm}[t]
\caption{Pseudocode for global residual vector assembly.}\label{alg:residual}
\begin{algorithmic}[1]
\Function{FormResidualVector}{$\mathbf{U}$}
  \State $\mathbf{U}_{\ell}\longleftarrow \mathbf{U}$
  \ForAll{element $e$ in subpatch}
    \State $\mathbf{U}_e\longleftarrow \mathbf{U}_\ell$
    \State $\mathbf{G}_e \longleftarrow \textsc{Geometry}(e)$
    \ForAll{quadrature point $q$ in element $e$}
      \State $\mathbf{x}_q,w_q \longleftarrow \textsc{Quadrature}(q)$
      \State $J_q,\left\{N_q\right\} = \textsc{ShapeFuns}(\mathbf{x}_q,\mathbf{G}_e)$
      \State $\mathbf{F}_q = \textsc{Integrand}(\mathbf{x}_q,\left\{N_q\right\},\mathbf{U}_e)$\label{alg:residual:integrand}
      \State $ J_qw_q\mathbf{F}_q\overset{+\ }{\longrightarrow} \mathbf{F}_e$
    \EndFor
    \State $\mathbf{F}_e\overset{+\ }{\longrightarrow} \mathbf{F}_{\text{cache}}$
  \EndFor
  \State $\mathbf{F}_{\text{cache}}\overset{+\ }{\longrightarrow} \mathbf{F}$
  \State return $\mathbf{F}$
\EndFunction
\end{algorithmic}
\end{algorithm}

\subsection{Numerical differentiation}\label{sec:num-diff}

In the process of setting up new codes to solve nonlinear problems,
one of the most time-consuming tasks involves the analytical
derivation and subsequent coding of Jacobians. While exploring new
ideas, models, and/or formulations to a problem, the expression of the
nonlinear residual may change considerably, triggering the need to
update the definition of the Jacobian. Moreover, the residual may
contain complicated expressions (e.g., non-trivial material
constitutive relations) whose derivatives are hard to
obtain. Numerical differentiation can alleviate these burdens by
providing a reliable way to approximate the Jacobian at the expense of
additional computational time.

PETSc has built-in facilities within its nonlinear solver
implementations to approximate Jacobians through numerical
differentiation using two different techniques.
\begin{itemize}
\item \emph{Matrix-free Newton--Krylov}. In this approach, a sparse
  matrix is never assembled. Instead, matrix-vector products involving
  the Jacobian are approximated within the Krylov linear solver using
  a first-order difference formula~\cite{Knoll2004}. Even though this
  method can considerably reduce the memory requirements as well as
  the computational time, the lack of an actual matrix prevents using
  black-box preconditioning techniques. Unless the user can provide a
  (matrix-free) problem-specific preconditioner, this methodology is
  only suitable for nonlinear problems where the Jacobian is well
  conditioned. Otherwise, linear solvers may fail to converge or
  require large iteration counts and thus, residual evaluations,
  leading to a drastic increase in computational time. Matrix-free
  techniques can be enabled at runtime through the command line option
  \texttt{-snes\_mf}.
\item \emph{Colored finite differences}. Unlike the previous approach,
  in this case a sparse matrix is assembled. A first-order difference
  formula is also used to form columns of the Jacobian matrix. Instead
  of computing columns individually, the method exploits sparsity by
  finding a partition of the set of columns using graph coloring
  techniques. In this way, many columns (more specifically, those
  having the same color) can be computed at once with a single global
  residual evaluation. We refer the reader to~\cite{Gebremedhin2005}
  for a thorough review on the topic. This black-box technique is
  sensible to use as long as the number of colors is small, which is
  the case in many discretization methods, and particularly linear
  finite elements. For higher-order finite element methods such as
  isogeometric analysis, the number of colors increases with the
  support of the basis functions. Colored finite differences
  techniques can be enabled at runtime through the command line option
  \texttt{-snes\_fd\_color}.
\end{itemize}

Being based on global residual evaluations, the approach
PETSc takes is black-box and applicable to many discretization methods
as it does not require explicit grid information. However, in the
context of finite element methods, it is possible to exploit locality
to develop an equivalent but more efficient implementation. Rather
than computing the global Jacobian by evaluating the global residual
vector, evaluations can instead be performed locally at the quadrature
point level.

In the following, we recall the notation in section~\ref{sec:assembly}
and details of algorithm~\ref{alg:residual}. In the local numerical
differentiation approach available in our framework, the $k$-th column
of the Jacobian matrix ${\partial\mathbf{F}_q}/{\partial\mathbf{U}_e}$
is computed at the quadrature point $q$ within an element $e$ using
the following first-order finite difference approximation
\begin{equation}\label{eq:fdjac}
  \dfrac{\partial\mathbf{F}_q}{\partial\mathbf{U}_e}\mathbf{\hat{e}}_k \approx
  \dfrac{1}{\delta}\Big(\mathbf{F}_q(\mathbf{U}_e+\delta\mathbf{\hat{e}}_k)-\mathbf{F}_q(\mathbf{U}_e)\Big)
\end{equation}
where $\mathbf{\hat{e}}_k$ denotes a vector with a $1$ in the $k$-th
entry and zeros elsewhere, while $\delta$ is the difference
step. Following the standard approach taken in~\cite{Pernice1998}, the
difference step is computed as
\begin{equation}
  \delta=\eta\sqrt{1+\|\mathbf{U}_e\|}
\end{equation}
where $\eta$ should be set to the square root of the relative error in
the evaluations of $\mathbf{F}_q$. As a default, this relative error
is set to $\eta=\sqrt{\epsilon}$, where $\epsilon$ is the floating
point machine epsilon (for double precision floating point numbers,
$\epsilon \approx 2.22 \times 10^{-16}$).

In appendix~\ref{app:num-diff}, we present benchmark
results comparing the numerical differentiation techniques mentioned
in this section when applied to the three-dimensional Bratu
equation. Even though explicitly coding the Jacobian is always the
fastest way of computing this matrix, the local approach discussed in
the previous paragraph seems to be a wiser choice when approximating
it numerically.

We consider numerical differentiation to be an extremely useful
feature to have at hand while prototyping and debugging. Ultimately,
users should weigh computational overhead against development effort
to decide whether numerical differentiation helps them achieve their
particular goals.

\subsection{Additional features}

Our framework uses standard tensor product Gauss--Legendre quadrature.
Tensor product Gauss--Lobatto quadrature rules are also available.
Research on quadrature rules tailored to isogeometric analysis started
in~\cite{Hughes2010}. Support for optimal, B-spline-specific,
quadrature rules is
ongoing~\cite{AitHaddou2014,Barton2015a,Barton2015b}. PetIGA also
supports tensor product basis functions defined as Lagrange
interpolants on either Newton--Cotes points (as in traditional finite
elements) or Gauss--Lobatto points (as in spectral finite
elements~\cite{Karniadakis2013}). Although these methods are not the
primary focus of our library, their availability can be useful to
researchers willing to explore isogeometric analysis as an alternative
to these well-established technologies. Even though PetIGA is geared
towards handling Galerkin-type finite element formulations, the
framework also supports isogeometric collocation
methods~\cite{Auricchio2010}.

\begin{figure}
\centering%
\subfloat{
\begin{minipage}[b]{0.45\textwidth}
\begin{lstlisting}[style=python,firstline=1,lastline=18]
from igakit.cad import *

C0 = circle(radius=1)
C1 = circle(radius=2)
annulus = ruled(C0, C1)

pipe = extrude(annulus,
               displ=3,
               axis=2,
               ).reverse(2)

elbow = revolve(annulus,
                point=(3,0,0),
                axis=(0,-1,0),
                angle=Pi/2)

bentpipe = join(pipe, elbow,
                axis=2)
\end{lstlisting}
\end{minipage}}
\subfloat{%
\includegraphics[width=0.4\textwidth]{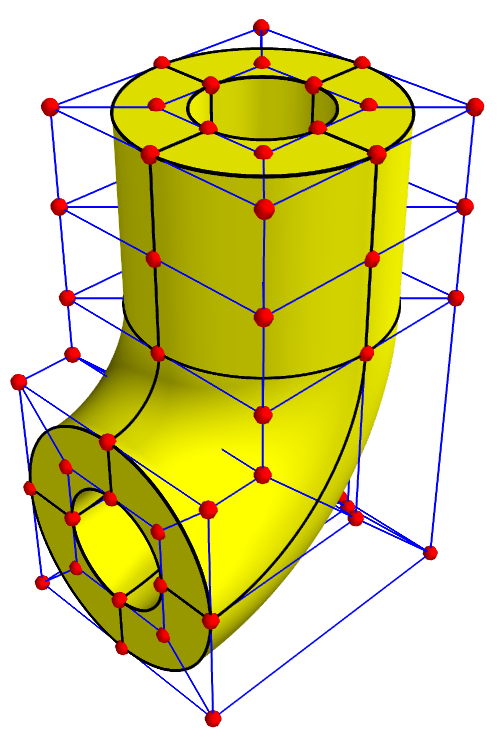}}
\caption{%
  Generating a bent pipe (see~\cite{Cottrell2009}, section~2.4) with
  igakit. The definition of this NURBS volume geometry entails four
  successive steps. First, an annulus is generated as a ruled surface
  between two concentric circles of different radii. Then, a pipe is
  generated by extruding the annulus along the appropriate
  direction. Next, an elbow is generated by revolving the base annulus
  $90^{\circ}$ around the appropriate axis. Finally, the pipe and the
  elbow are joined along one of the parametric directions.}
\label{fig:bentpipe}
\end{figure}

Creation of initial geometries is a nontrivial task. Representation of
a volume using NURBS is cumbersome and mostly developed manually, and
bridging Computer Aided Design and Computer Aided Engineering is
ongoing. To ease the handling of NURBS geometries we developed
igakit~\cite{igakit}, a Python package providing low-level interfaces
to Fortran 90 routines in charge of manipulating knot vectors and
control point data, as well as a high-level interface to simplify the
definition of these geometries by hand. Details on using igakit to
create geometries such as the one presented in
figure~\ref{fig:bentpipe} can be found in~\cite{igakit-tutorial}.

\lstdefinestyle{python}{
  language=Python,
  frame=L,
  lineskip={1pt},
  xleftmargin=\parindent,
  numbers=left,
  basicstyle=\footnotesize\ttfamily,
  numberstyle=\tiny\ttfamily,
}

\section{Example}\label{sec:example}
In this section we solve the Bratu equation as a model application to
highlight some of the useful features users have access to when using
our framework.

\subsection{The Bratu equation}

The Bratu equation is a nonlinear second-order boundary value
problem~\cite{Jacobsen2002}. The strong form of the equation can be
stated as: find $u: \bar{\Omega}\rightarrow\mathbb{R}$ such that
\begin{alignat}{4}
  -\Delta u &= \lambda \exp(u),\quad && \boldsymbol{x} \ &&\in \ &&\Omega,\label{eq:strong}\\
  u &= 0, && \boldsymbol{x} &&\in &&\partial \Omega\label{eq:dirichlet},
\end{alignat}
where $\bar{\Omega}$ is the unit square in $[0,1]^2$,
$\partial \Omega$ denotes the domain boundary, $\Delta$ represents the
Laplace operator, $u \equiv u(\boldsymbol{x})$ is a scalar field defined
in $\Omega$ and $\lambda$ is a positive constant. No solution exists
when $\lambda$ goes above $\lambda_{max}=6.80812$ as the equation
possesses a bifurcation point. This equation models the steady-state
of a nonlinear reaction and heat conduction problem, and results from
a simplification of the solid-fuel ignition model.

Within the framework of Galerkin finite elements, let $\mathcal{V}$
denote the trial and weighting function spaces, where $\mathcal{V}$
belongs to $\mathcal{H}^1_0$, i.e., the Sobolev space of
square-integrable functions with square-integrable first derivatives
and zero value on $\partial\Omega$. The weak form is obtained by
multiplying equation~\eqref{eq:strong} by a test function $w$ and
integrating by parts. The variational problem can then be defined as
that of finding $u \in \mathcal{V}$ such that for all
$w \in \mathcal{V}$
\begin{align}
  \left(\nabla w, \nabla u\right)_{\Omega} - \left(w,\lambda \exp\left(u\right)\right)_{\Omega}=0,\label{eq:wf}
\end{align}
where $\left(\cdot,\cdot\right)_{\Omega}$ denotes the $\mathcal{L}^2$
inner product on domain $\Omega$. The finite-dimensional problem can
then be formulated as: find $u^h \in \mathcal{V}^h$, where
$\mathcal{V}^h \subset \mathcal{V}$, such that for all
$w^h \in \mathcal{V}^h$
\begin{align}
  \left(\nabla w^h, \nabla u^h\right)_{\Omega} - \left(w^h,\lambda \exp\left(u^h\right)\right)_{\Omega}=0,\label{eq:wfd}
\end{align}
where $w^h$, $u^h$ and their respective gradients are defined as the
linear combinations
\begin{align}
  w^h        &= \sum_{A}W_AN_A(\boldsymbol{x}),          & u^h        &= \sum_{A}U_AN_A(\boldsymbol{x}),         \label{eq:wuh0} \\
  \nabla w^h &= \sum_{A} W_A \nabla N_A(\boldsymbol{x}), & \nabla u^h &= \sum_{A}U_A \nabla N_A(\boldsymbol{x}), \label{eq:wuh1}
\end{align}
where $N_A$ are the basis functions and $W_A$, $U_A$ are the control
variables. Denoting $\mathbf{U}=\left\{U_A\right\}$ the vector of
coefficients, we define the residual vector as
$\mathbf{R\left(U\right)}=\left\{R_A\right\}$, where $R_A$ is obtained
from~\eqref{eq:wfd}--\eqref{eq:wuh1} as
\begin{align}
  R_A=\left(\nabla N_A, \nabla u^h\right)_{\Omega} - \left(N_A,\lambda \exp\left(u^h\right)\right)_{\Omega},\label{eq:residual}
\end{align}

As the problem is nonlinear, solving it with Newton's method requires
the specification of the Jacobian
$\mathbf{J} = \partial\mathbf{R}/\partial\mathbf{U}$. In this
particular problem, its entries are defined as
\begin{align}
  J_{AB} = \left(\nabla N_A,\nabla N_B\right)_{\Omega} - \left(N_A,\lambda\exp\left(u^h\right) N_B\right)_{\Omega}.\label{eq:jacobian}
\end{align}

\subsection{Implementation}

\lstdefinestyle{C}{
  language=C,
  frame=L,
  lineskip={1pt},
  xleftmargin=\parindent,
  numbers=left,
  basicstyle=\footnotesize\ttfamily,
  numberstyle=\tiny\ttfamily,
  keywordstyle=\bfseries\color{green!40!black},
  morekeywords={typeof},
}

\newcommand{\codelisting}[4][]{
\lstinputlisting[style=C,firstline=#2,lastline=#3,firstnumber=#2,#1]{#4}
}

To code a program within the framework, the user first needs to
include the PetIGA C header file%
\codelisting{1}{1}{lst-Bratu.c}%
For the sake of simplicity, two macros are defined to set the
dimensionality of the problem to two and implement the dot product of
two-vectors%
\codelisting{3}{4}{lst-Bratu.c}%
To change the dimensionality of the problem to one or three, these two
lines should be modified accordingly. Then, a C structure is used to
handle the problem-specific parameter $\lambda$. This approach is
preferred to the alternative of using global variables.%
\codelisting{6}{8}{lst-Bratu.c}%

The residual routine implementing equation~\eqref{eq:residual} reads%
\codelisting{10}{20}{lst-Bratu.c}%
Recalling section~\ref{sec:assembly}, the residual routine constitutes
the integrand to be evaluated at each quadrature point
(line~\ref{alg:residual:integrand} of algorithm~\ref{alg:residual}) to
compute local contributions to the global residual vector. The
\texttt{Residual} routine has the following arguments
\begin{itemize}
\item an input pointer \texttt{p}, of type \texttt{IGAPoint}, used as
  a quadrature point context holding discretization data required to
  perform the residual evaluation,
\item an input floating point array \texttt{U}, which contains the
  local control variables gathered from the global vector
  $\mathbf{U}$, i.e., the coefficients corresponding to basis
  functions whose support contains the quadrature point, recall
  equations~\eqref{eq:wuh0}~and~\eqref{eq:wuh1},
\item an output floating point array \texttt{R}, where the routine is
  expected to return local contributions to be assembled in the global
  residual vector $\mathbf{R}$, recall equation~\eqref{eq:residual},
\item an input opaque pointer \texttt{ctx}, used to pass
  problem-specific information to the residual routine.
\end{itemize}
The code proceeds to declare the following local variables
\begin{itemize}
\item two integers \texttt{a} and \texttt{nen}, the first to be used
  as a loop index, while the second is initialized to the number of
  local basis functions,
\item two pointers \texttt{N0} and \texttt{N1}, initialized from data
  within the \texttt{IGAPoint}, used in the following for convenient
  access to the values of local basis functions ($0$th derivatives)
  and their first derivatives,
\item two floating point variables \texttt{u} and \texttt{grad\_u} to
  store the values of $u^h$ and $\nabla u^h$ at the quadrature point
\item a floating point variable \texttt{lambda} to store the value of
  $\lambda$, initialized from the \texttt{Params} structure through
  the opaque pointer \texttt{ctx}.
\end{itemize}
Next, the routines \texttt{IGAPointFormValue} and
\texttt{IGAPointFormGrad} are invoked to compute the values of $u^h$
and $\nabla u^h$ at the quadrature point, following
equations~\eqref{eq:wuh0}~and~\eqref{eq:wuh1}.
Finally, local residual contributions are computed in a loop following
equation~\eqref{eq:residual} by using the previously defined
\texttt{dot} macro as well as the \texttt{exp} routine from the
standard library of the C programming language.  The quadrature
weights and Jacobian determinant of the geometry mapping are handled
internally, which slightly simplifies the coding.

The Jacobian routine implementing equation~\eqref{eq:jacobian} reads%
\codelisting{22}{32}{lst-Bratu.c}%
The \texttt{Jacobian} routine is almost the same as the previous
\texttt{Residual} routine. The main differences are the output
floating point array \texttt{J} where the routine is expected to
return local contributions to be assembled in the global Jacobian
matrix $\mathbf{J}$, and the double-loop computing these contributions
following equation~\eqref{eq:jacobian}.

The code of the main program routine then begins%
\codelisting{34}{35}{lst-Bratu.c}%
The first statement invokes the \texttt{PetscInitialize} routine,
which internally handles the initialization of the PETSc library. This
routine is fed with the command line arguments to the program. PETSc
uses these arguments to build a database of options. These options are
queried later.

The code proceeds to create and initialize an \texttt{iga} object of
type \texttt{IGA}. The \texttt{IGA} type is a key component in
PetIGA. This core data structure contains all the information related
to discretization and parallel communication.%
\codelisting{37}{42}{lst-Bratu.c}%
To properly setup the \texttt{iga} object, the user only needs to
hardwire in code a couple of problem-specific parameters, and ask the
framework to handle the rest at runtime through the options database
\begin{itemize}
\item the routine \texttt{IGASetDim} specifies the number of space
  dimensions, in this particular example it is set to two,
\item the routine \texttt{IGASetDof} specifies the number of
  components in the solution, in this particular example it is set to
  one as the code is dealing with a scalar problem,
\item the routine \texttt{IGASetFromOptions} queries the options
  database to let users change different properties of the
  discretization such as number of elements, polynomial degree and
  regularity of the approximation space, type of basis functions to
  use, quadrature rules, number of quadrature points, among
  others. When values are not specified by the user, PetIGA selects
  default values such as $16$ elements per direction, unit domain,
  uniform refinement, quadratic $C^1$ spaces, and Gauss--Legendre
  quadrature. Even though other ways to initialize an IGA object are
  available, this is the simplest one. When more complex
  discretizations and/or geometries are required, the IGA object can
  be initialized by loading binary datafiles,
\item finally, the routine \texttt{IGASetUp} prepares the data
  structure to be used in what follows. This step handles the parallel
  partitioning of the domain and prepares internal data structures
  that manage parallel communication.
\end{itemize}

Boundary conditions are handled in the code snippet that follows%
\codelisting{44}{50}{lst-Bratu.c}%
The routine \texttt{IGASetBoundaryValue} is used to set the Dirichlet
boundary conditions of this problem, as specified in
equation~\eqref{eq:dirichlet}, and takes as arguments
\begin{itemize}
\item the \texttt{IGA} context in which the boundary condition is to
  be set,
\item the parametric direction, which takes values \texttt{0} or
  \texttt{1} to specify either the first or second parametric
  directions, respectively,
\item the boundary side, which takes values \texttt{0} or \texttt{1}
  to specify either the left or right side along the parametric
  direction, respectively,
\item the component index, which is \texttt{0} for scalar problems,
\item finally, the value to be enforced at the boundary.
\end{itemize}
Although not highlighted in this example, PetIGA allows users to
specify more general and possibly nonlinear boundary conditions. These
boundary conditions should be handled in the user-defined residual and
Jacobian routines.

The final step to configure the \texttt{IGA} context requires the
specification of the user-defined residual and Jacobian callbacks, as
shown in the following lines of code%
\codelisting{52}{55}{lst-Bratu.c}%
A local variable \texttt{params} of type \texttt{Params} is declared,
the \texttt{lambda} member is initialized from a user-specified
command line option, or from a hardwired default value otherwise. The
calls to \texttt{IGASetFormFunction} and \texttt{IGASetFormJacobian}
register within the \texttt{IGA} context the user-defined residual and
Jacobian routines. They also store the pointer to the \texttt{Params}
instance which is used in \texttt{Residual} and \texttt{Jacobian} to
access the $\lambda$ parameter, as explained previously.

The code then proceeds to solve the nonlinear problem%
\codelisting{57}{63}{lst-Bratu.c}%
The routine \texttt{IGACreateSNES} creates a nonlinear solver context
and performs additional initialization such as associating the global
vector to form the residual and the global matrix in which to form the
Jacobian within the solver context. The routine
\texttt{SNESSetFromOptions} enables users to further configure the
nonlinear solver through command line options. Finally, a global vector
is created to store the solution coefficients and the routine
\texttt{SNESSolve} is invoked to solve the problem.

A call to the routine~\texttt{VecViewFromOptions} creates a viewer
context which can be used to visualize the solution vector in
real-time, or store the solution as a VTK
datafile~\cite{vtk-book,vtk-file-formats}%
\codelisting{65}{65}{lst-Bratu.c}%

Lastly, all PETSc and PetIGA objects created through the code are
destroyed to free resources and the routine \texttt{PetscFinalize} is
invoked to properly shutdown the framework%
\codelisting{67}{72}{lst-Bratu.c}%


\section{Applications}\label{sec:applications}
To illustrate the flexibility of our software, we showcase
applications which highlight strengths of isogeometric analysis. The
problems come from standard engineering domains, such as solid and
fluid mechanics, as well as from less traditional areas, such as
phase-field modeling. We have chosen to solve nonlinear, partial
differential equations in order to demonstrate our code framework on a
challenging subset of problems. In each problem, a nonlinear residual
functional is obtained from the weak form of the partial differential
equation. Many problem-specific details we omit here and refer the
reader to the referenced software package~\cite{petiga} where they can
find full details in the form of demonstration programs.

\subsection{Steady-state Hyper-elasticity}

The first of our examples is a hyper-elastic material model in the
context of large deformation, applied to a cylindrical
tube~\cite{Bernal2013}. The strong form can be expressed as: find the
displacement $\mathbf{U}:\overline{\Omega}\mapsto\mathbb{R}^3$ such
that
\begin{alignat*}{3}
  \nabla\cdot\mathbf{P} &= \mathbf{0}\quad &&\text{in}\quad &&\Omega,\\
  \mathbf{U} &= \mathbf{G} &&\text{on} &&\Gamma_D,\\
  \mathbf{P}\cdot\mathbf{N} &= \mathbf{0} &&\text{on} &&\Gamma_N,
\end{alignat*}
where $\mathbf{P}$ is the first Piola--Kirchhoff stress tensor,
$\mathbf{G}$ is the prescribed displacement on the Dirichlet boundary
$\Gamma_D$, $\mathbf{N}$ is the outward normal of the Neumann boundary
$\Gamma_N$ in the reference configuration
(see~\cite{SimoHughes,ContMech} for more details). We use a
Neo-Hookean material model, which relates the second Piola--Kirchhoff
stress tensor $\mathbf{S}$ to the right Cauchy--Green strain tensor
$\mathbf{C}$ by the relationship
\begin{equation*}
  \mathbf{S} = 
  \frac{\lambda}{2}\left(J^2-1\right)\mathbf{C}^{-1} + 
  \mu\left(\mathbf{I}-\mathbf{C}^{-1}\right),
\end{equation*}
where $\mathbf{C}=\mathbf{F}^\mathrm{T}\mathbf{F}$, $\mathbf{F}$ is
the deformation gradient, $J=\det(\mathbf{F})$, and $\lambda,\mu$ are
the Lam\'e constants from linear elasticity. The first
Piola--Kirchhoff stress tensor is defined by
$\mathbf{P}=\mathbf{F}\mathbf{S}$ and the symmetry condition
$\mathbf{P}^\mathrm{T}\mathbf{F}=\mathbf{F}^\mathrm{T}\mathbf{P}$.

\begin{figure}
\centering
\subfloat[Undeformed tube] {\includegraphics[width=0.48\textwidth]{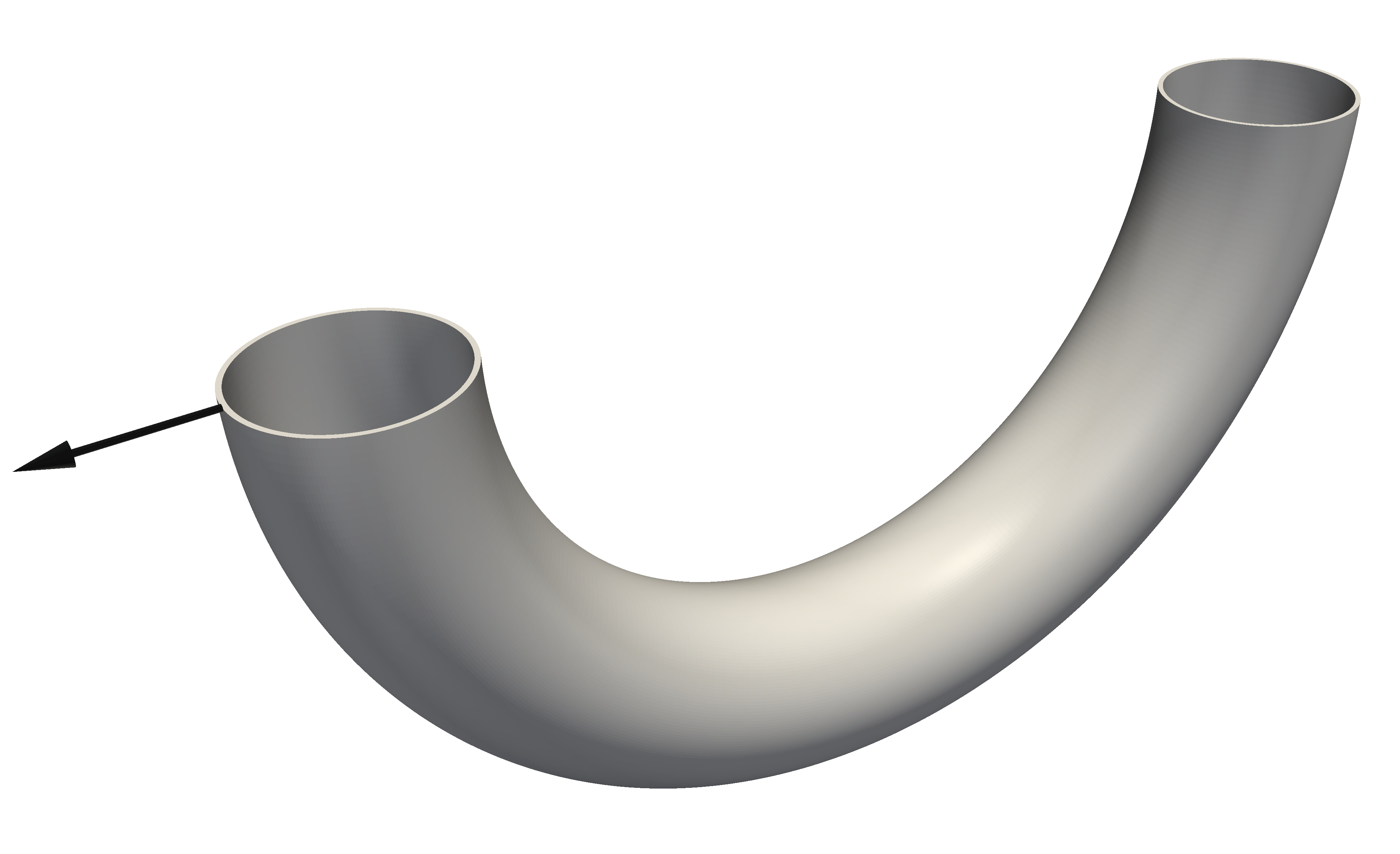}}
\subfloat[Deformed tube]   {\includegraphics[width=0.48\textwidth]{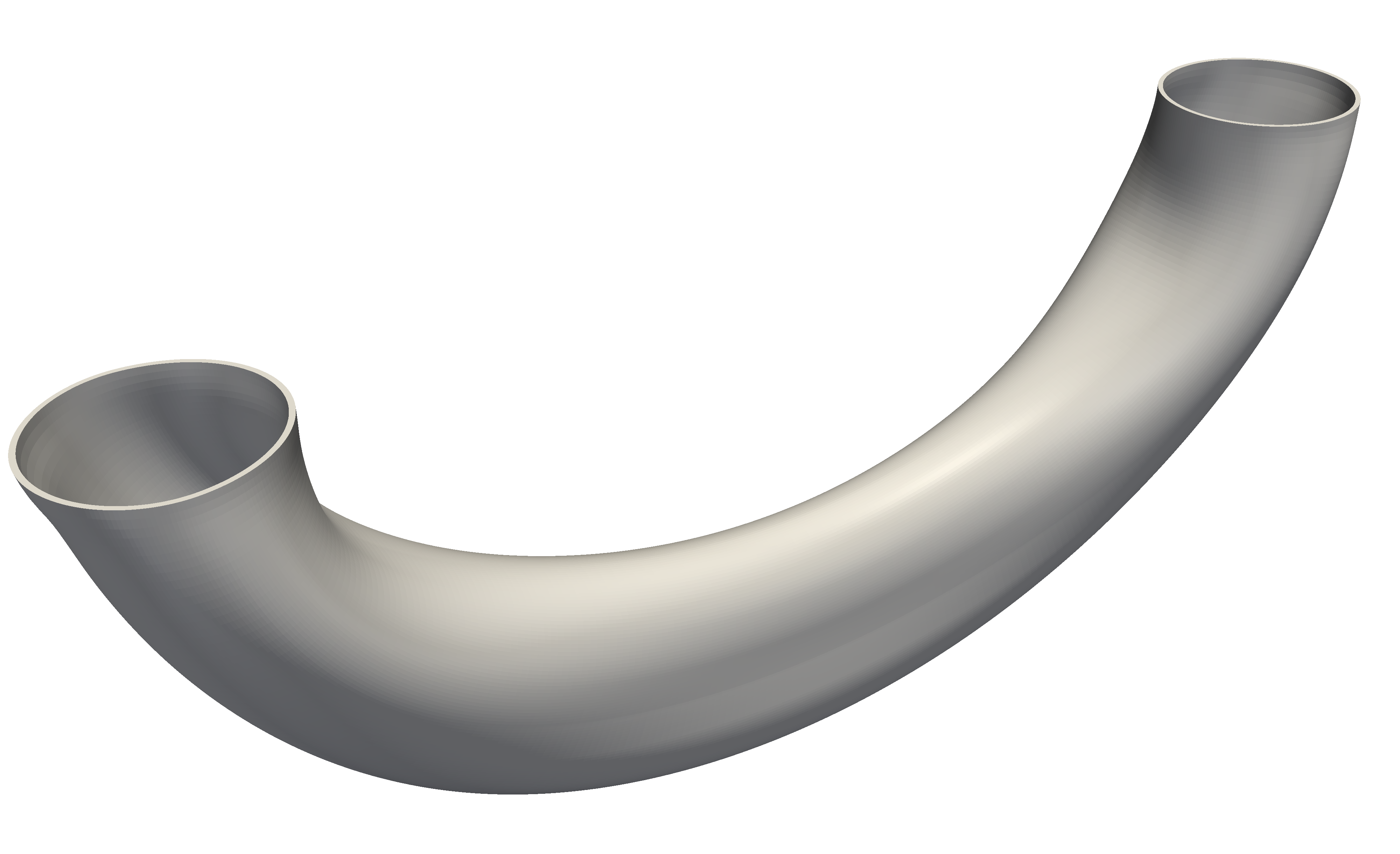}}
\caption{%
  Deformation of an aluminum tube, modeled using a Neo-Hookean
  material model. The arrow represent the direction and magnitude of
  the displacement imposed as a boundary condition on the left end
  while the right end is kept fixed.}
\label{fig:tube}
\end{figure}

We solve the linearized weak form of these equations using Newton's
method in an updated-Lagrangian approach. In figure~\ref{fig:tube}(a)
we depict the reference geometry, a circular tube discretized with a
mesh of $16 \times 64 \times 4$ quadratic B-spline functions. The
right side of the tube is fixed and the left side is displaced to the
left over 15 load steps. In this case we configured PETSc to interface
with MUMPS and solved the system using this parallel direct
solver. The deformed configuration is shown in
figure~\ref{fig:tube}(b).

\subsection{Time-dependent Problems}

The following three examples discretize time-dependent, nonlinear
problems. We first detail our solution strategy for these
problems. For simplicity, we consider a scalar problem. We seek to
find $\dot{u},u \in \mathcal{U}$ such that
\begin{equation*}
  \mathcal{R}(w;\dot{u},u) = 0\quad \forall w \in \mathcal{W},
\end{equation*}
where $\mathcal{R}$ is a time-dependent, nonlinear residual functional.

We first use a semi-discrete approach by discretizing $\dot{u},u$ in
space with finite-dimensional subspaces $\mathcal{U}^h \subset
\mathcal{U}$, leaving the problem continuous in time. The span of the
set of basis functions $\{N_B(\boldsymbol{x})\}_{B=0\dots n}$,
$\boldsymbol{x} \in \Omega$ define the subspace
$\mathcal{U}^h$. Similarly, we choose $\mathcal{W}^h \subset
\mathcal{W}$ where $\text{span}\left(\{N_A(\boldsymbol{x})\}_{A=0\dots
    n-1}\right)$ defines $\mathcal{W}^h$. Given two sequences of
time-dependent coefficients $\{\dot{U}_B(t)\},\{U_B(t)\}$, discrete
functions $\dot{u}^h,u^h$ are defined as
\begin{align*}
  \dot{u}^h(\boldsymbol{x},t) &= \sum_B \dot{U}_B(t)N_B(\boldsymbol{x}),\\
  u^h(\boldsymbol{x},t) &= \sum_B U_B(t)N_B(\boldsymbol{x}).
\end{align*}
Denoting $\mathbf{\dot{U}}=\{\dot{U}_B(t)\},\mathbf{U}=\{U_B(t)\}$, we
can write the residual vector,
\begin{equation*}
  \mathbf{R}\left(\mathbf{\dot{U}},\mathbf{U}\right) = \{R_A\},
\end{equation*}
where
\begin{equation*}
  R_A = \mathcal{R}\left(N_A;\dot{u}^h,u^h\right).
\end{equation*}

Next, we discretize in time using the generalized-$\alpha$ method for
first order systems~\cite{Jansen2000}. Given
$\mathbf{\dot{U}}_n,\mathbf{U}_n$, we seek
$\mathbf{\dot{U}}_{n+1},\mathbf{U}_{n+1}$ such that
\begin{align}\label{eq:ga}
  \mathbf{R}\left(\dot{\mathbf{U}}_{n+\alpha_m},\mathbf{U}_{n+\alpha_f}\right) & =0\\
  \mathbf{\dot{U}}_{n+\alpha_m} &= \mathbf{\dot{U}}_{n}+\alpha_m\left(\mathbf{\dot{U}}_{n+1}-\mathbf{\dot{U}}_{n}\right)\\
  \mathbf{U}_{n+\alpha_f} &= \mathbf{U}_{n}+\alpha_f\left(\mathbf{U}_{n+1}-\mathbf{U}_{n}\right)\\
  \mathbf{U}_{n+1} &= \mathbf{U}_n + \Delta t\left(\gamma \dot{\mathbf{U}}_{n+1} + (1-\gamma)\dot{\mathbf{U}}_{n} \right)
\end{align}
where $\Delta t=t_{n+1}-t_{n}$ is the time step, and
$\alpha_f,\alpha_m,\gamma$ are parameters which define the method. The
generalized-$\alpha$ method was designed to filter (damp)
high-frequency modes of the solution which are under-approximated. As
opposed to linear problems, low and high frequency modes interact in
nonlinear problems. Spurious high frequency modes lead to
contamination of the resolved modes of the problem. The method
parameters $\alpha_f,\alpha_m,\gamma$ can be chosen using the spectral
radius $\rho_{\infty} \in [0,1]$ of the amplification matrix as
$\Delta t\rightarrow\infty$ by
\begin{align*}
  \alpha_m &= \frac{1}{2}\left(\frac{3-\rho_{\infty}}{1+\rho_{\infty}}\right)\\
  \alpha_f &= \frac{1}{1+\rho_{\infty}}\\
  \gamma & = \frac{1}{2} + \alpha_m - \alpha_f
\end{align*}
which leads to a second order, unconditionally stable method. The
value of $\rho_{\infty}$ uniquely defines the method and can be chosen
to filter a desired amount of high frequency modes.

In our experience, when approaching time-dependent,
nonlinear problems, the generalized-$\alpha$ method is effective. The
impact of the interaction of low and high frequency modes is problem
dependent and not something necessarily understood in advance. In the
case where no filtering is needed ($\rho_{\infty}=1$), the
generalized-$\alpha$ method reduces to the implicit midpoint rule. Due
to the popularity of the generalized-$\alpha$ method particularly
among researchers in the isogeometric community, we have added it to
PETSc's time stepping algorithms and is available independently of the
PetIGA framework. In the examples that follow, we solve
equation~\eqref{eq:ga} iteratively using Newton's method, which
requires the computation of the Jacobian of the residual vector at
each iteration.

\subsubsection{Cahn--Hilliard Equation}

The Cahn--Hilliard equation governs the evolution of a binary mixture
undergoing the process of phase separation. Let $\Omega \in
\mathbb{R}^d$ be an open set, where $d=2,3$. The boundary of $\Omega$
with unit outward normal $\boldsymbol{n}$ is denoted $\Gamma$ and is
composed of two complementary parts $\Gamma_g$ and
$\Gamma_h$. Denoting $c$ the concentration of one of the components of
the mixture, the problem can be stated in strong form as: find $c :
\overline{\Omega} \times (0,T) \mapsto \mathbb{R}$ such that
\begin{alignat*}{3}
  \dfrac{\partial c}{\partial t} - \nabla\cdot\left(M_c\nabla\left(\mu_c-\lambda\Delta{c}\right)\right) &= 0\quad &&\text{in}\quad &&\Omega \times (0,T),\\
  c &= g &&\text{on} &&\Gamma_g \times (0,T),\\
  M_c \nabla\left(\mu_c-\lambda\Delta{c}\right)\cdot \boldsymbol{n} &= h  &&\text{on} &&\Gamma_h \times (0,T),\\
  M_c \lambda \nabla c \cdot \boldsymbol{n} &= 0 &&\text{on} &&\Gamma \times (0,T),\\
  c\left(\boldsymbol{x},0\right) &= c_0\left(\boldsymbol{x}\right) \quad &&\text{in} &&\overline{\Omega},
\end{alignat*}
where $c_0$ is the initial concentration, $M_c$ is the mobility,
$\mu_c$ represents the chemical potential of a binary mixture in the
absence of phase interfaces, and $\lambda$ is a positive constant such
that $\sqrt{\lambda}$ represents a length scale of the problem related
to the interface thickness between the two phases.  The mobility and
chemical potential are nonlinear functions of the concentration
\begin{align*}
  M_c &= Dc(1-c),\\
  \mu_c &= \dfrac{1}{2\theta}\log\dfrac{c}{1-c}+1-2c,
\end{align*}
in which $D$ is a positive constant with dimensions of diffusivity and
$\theta$ is a dimensionless number which represents the ratio between
critical and absolute temperatures.

\begin{figure}[h]
  \centering
  \includegraphics[width=0.35\textwidth]{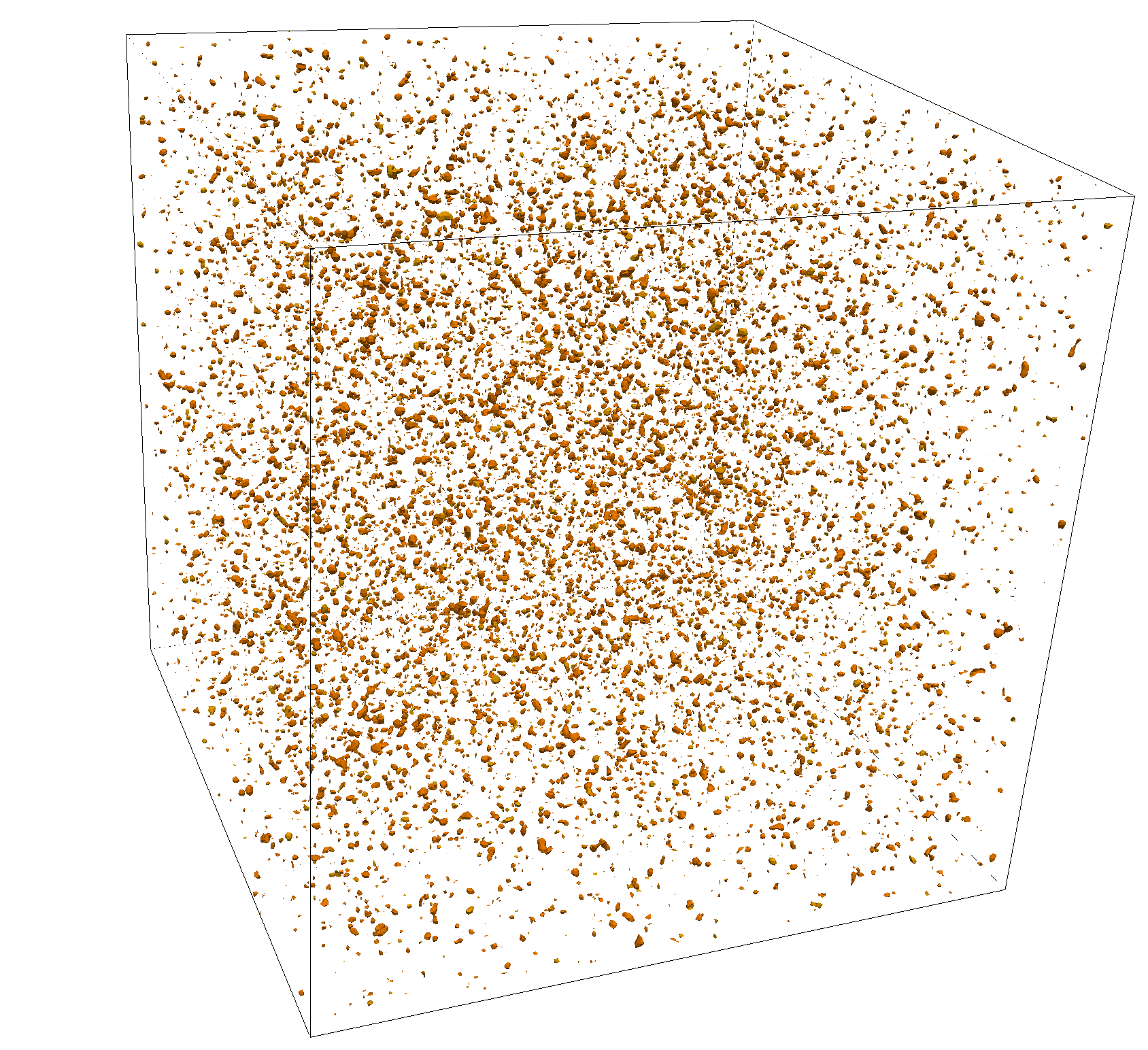}\qquad
  \includegraphics[width=0.35\textwidth]{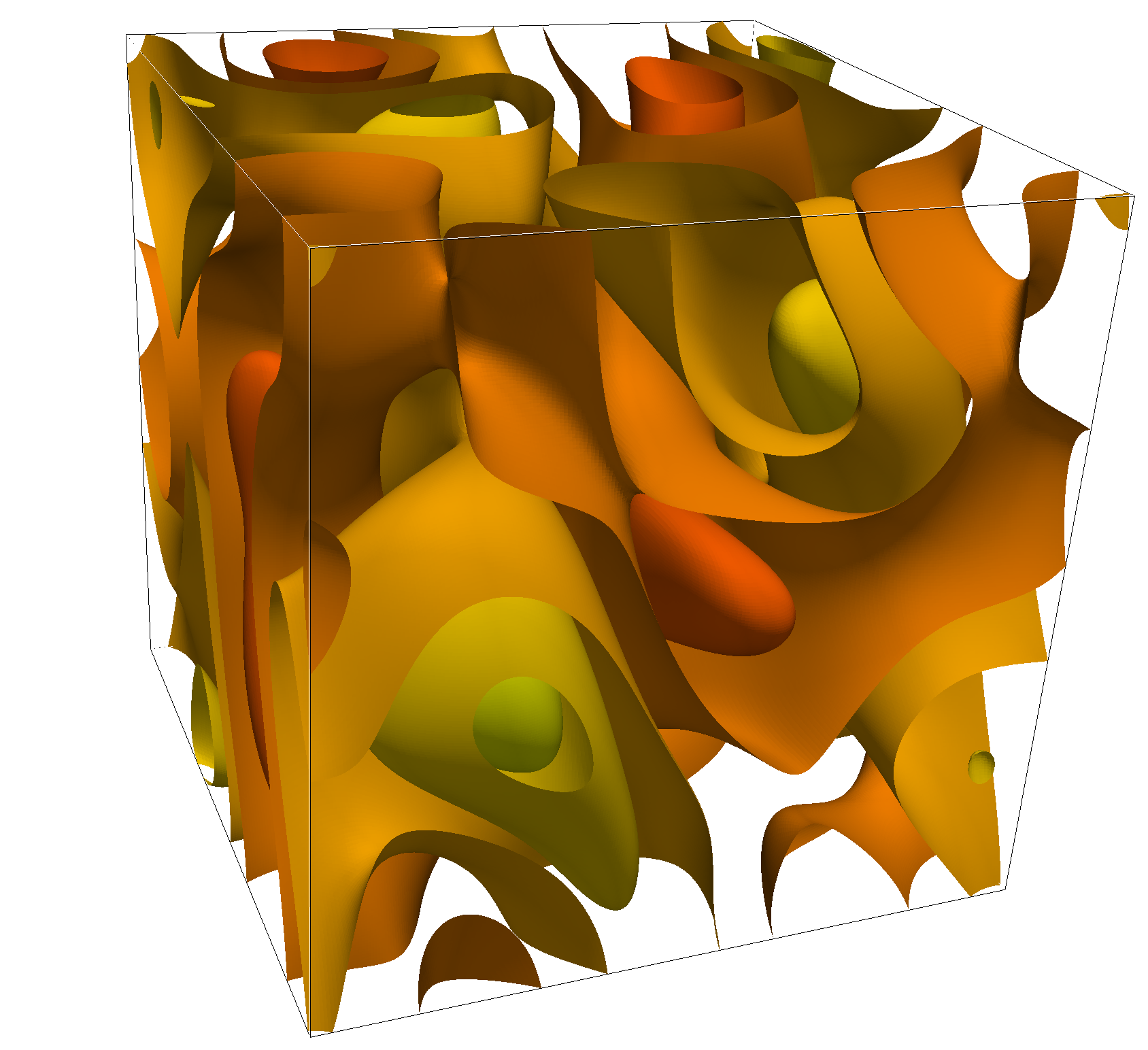}\\
  \includegraphics[width=0.35\textwidth]{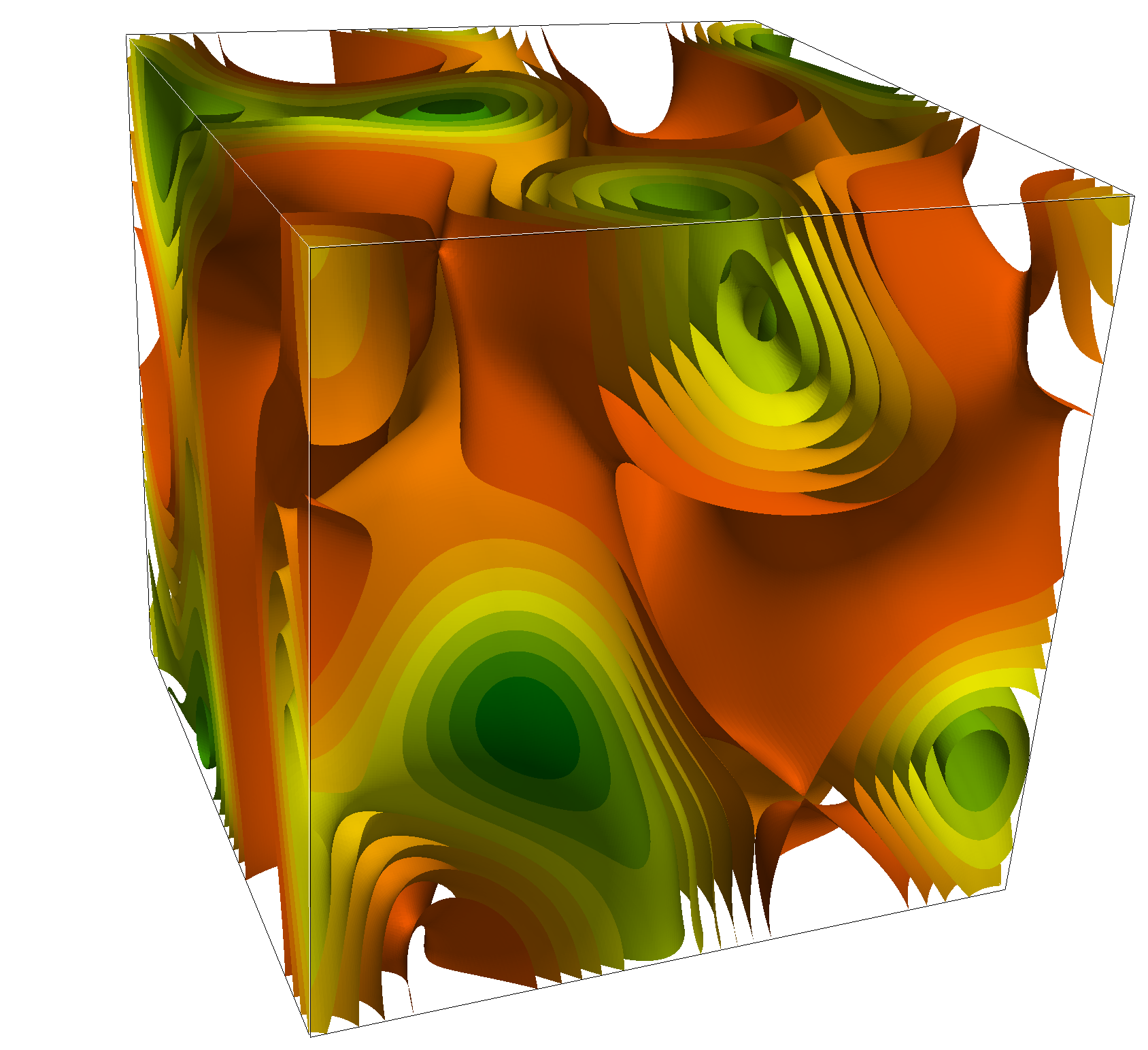}\qquad
  \includegraphics[width=0.35\textwidth]{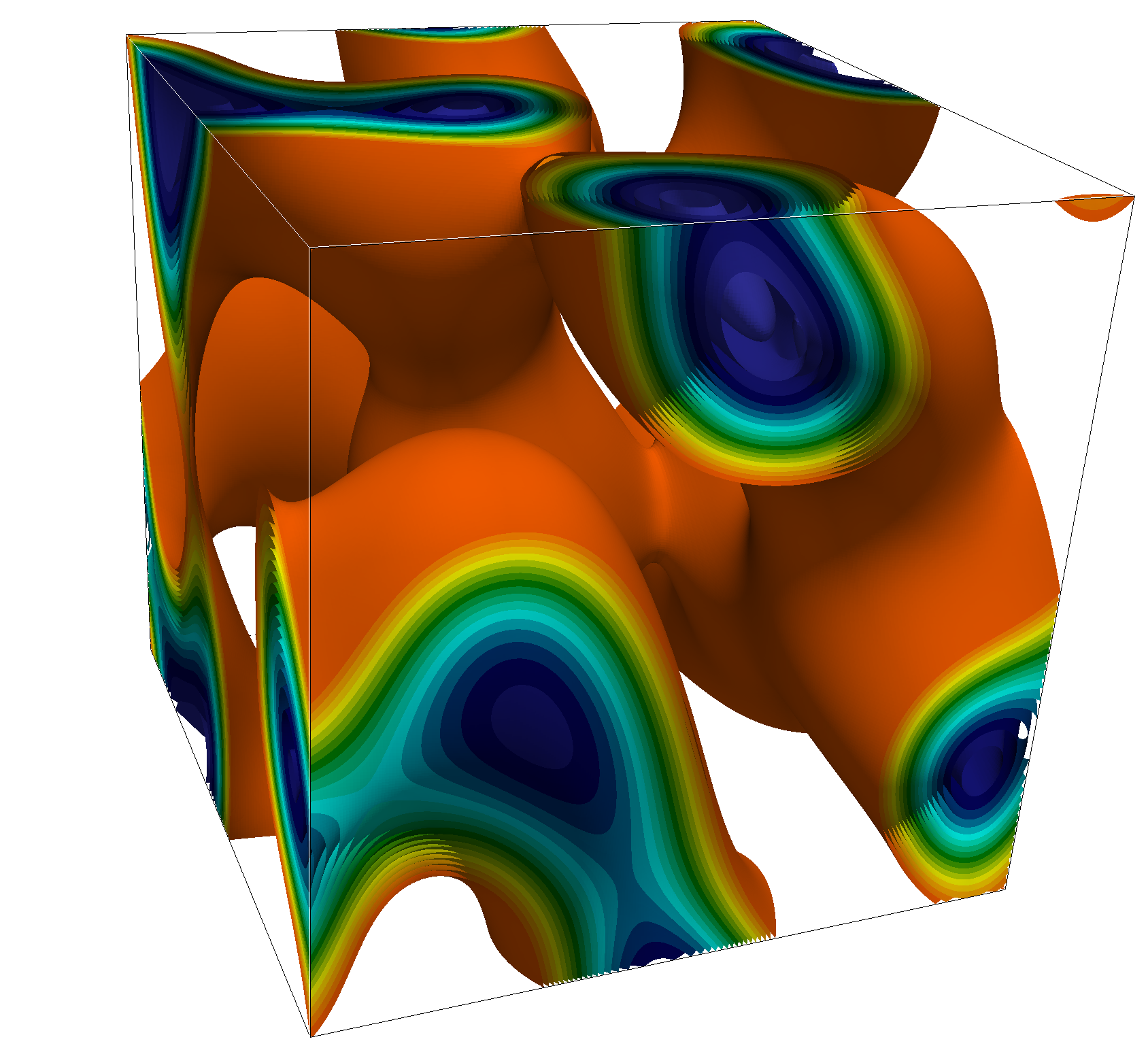}\\
  \includegraphics[width=0.35\textwidth]{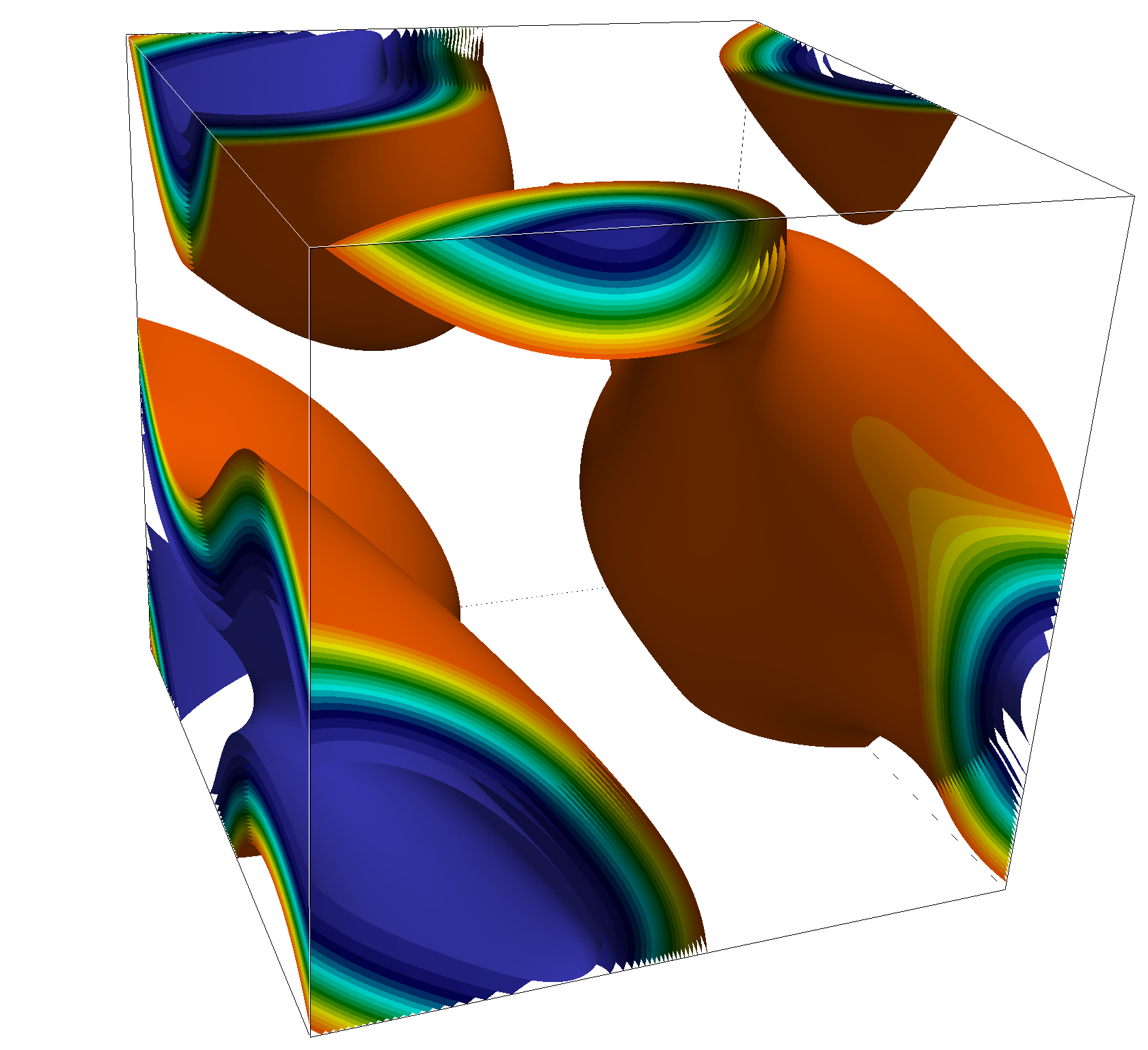}\qquad
  \includegraphics[width=0.35\textwidth]{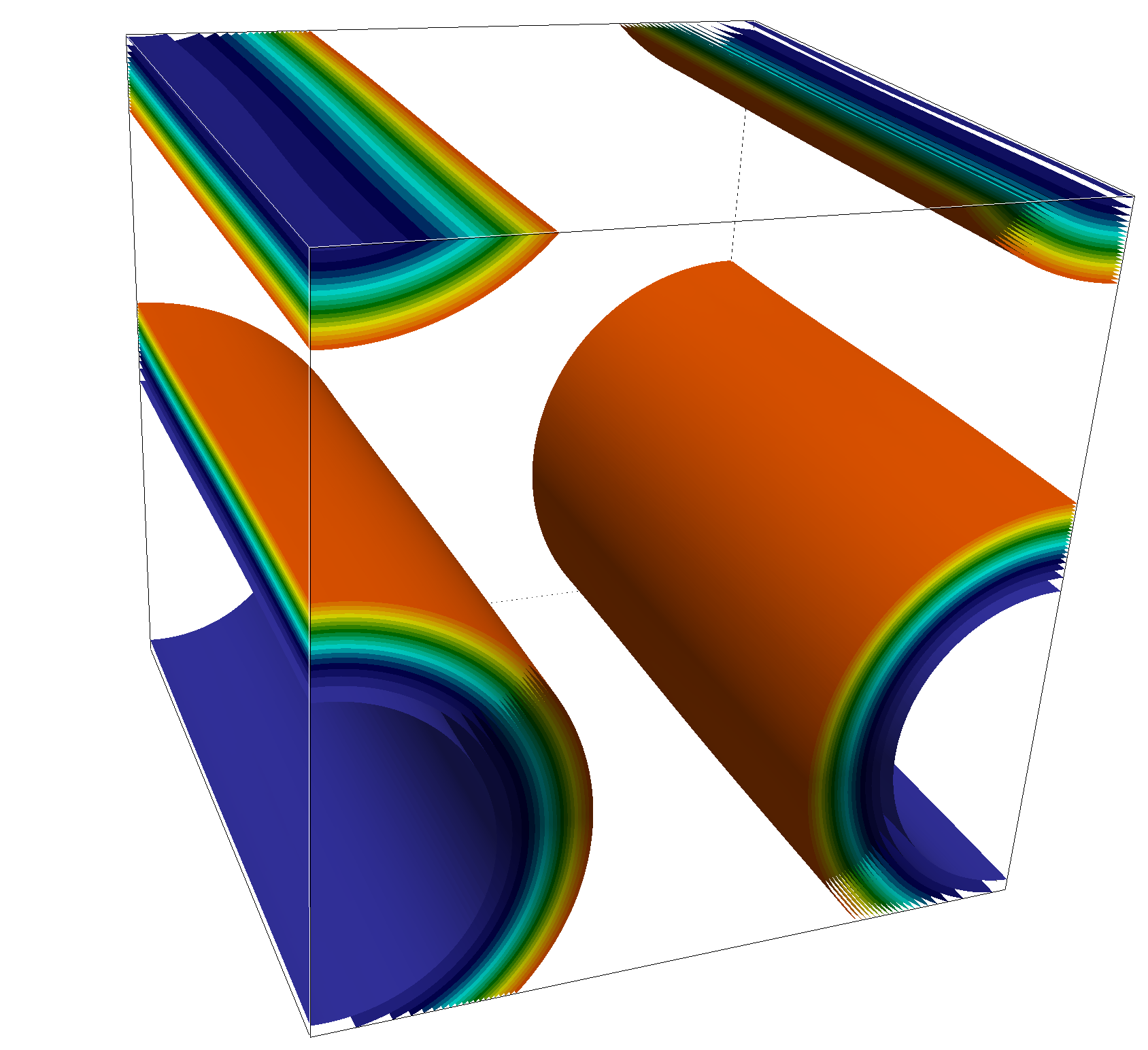}
  \caption{%
    Transient solution to the Cahn--Hilliard problem in three
    dimensions, subject to a random initial condition and periodic
    boundary conditions. The weak form is discretized in space by a mesh
    of $256^3$ elements of $C^1$ quadratic B-splines.}
  \label{fig:ch}
\end{figure}

We solve the Cahn--Hilliard equation in dimensionless form and with
periodic boundary conditions as detailed in \cite{Gomez2008}. We show
snapshots of the isocontours of the solution in three dimensions at
different times during the simulation in figure~\ref{fig:ch}.

\subsubsection{Navier--Stokes--Korteweg Equations}

The Navier--Stokes--Korteweg equations are a phase-field model for
water and water vapor two-phase flows. Let $\Omega \in \mathbb{R}^d$
be an open set, where $d=2,3$. The boundary of $\Omega$ with unit
outward normal $\boldsymbol{n}$ is denoted $\Gamma$. The problem can
be stated in strong form as: find the density
$\rho:\overline{\Omega}\times(0,T)\mapsto(0,b)$ and the velocity
$\boldsymbol{u}:\overline{\Omega}\times(0,T)\mapsto\mathbb{R}^d$ such
that
\begin{alignat*}{3}
  \dfrac{\partial \rho}{\partial t} + \nabla \cdot \left( \rho\boldsymbol{u}\right) &= 0\quad &&\text{in}\quad &&\Omega \times (0,T),\\
  \dfrac{\partial\left(\rho\boldsymbol{u}\right)}{\partial t} + \nabla\cdot\left(\rho\boldsymbol{u}\otimes\boldsymbol{u}+p\boldsymbol{I}\right)
  -\nabla\cdot\boldsymbol{\tau} - \nabla\cdot\boldsymbol{\zeta} &= \rho\boldsymbol{f}\quad &&\text{in} &&\Omega \times (0,T),\\
  \boldsymbol{u} &= \boldsymbol{0}\quad &&\text{on} &&\Gamma \times (0,T),\\
  \nabla\rho\cdot\boldsymbol{n} &= 0\quad &&\text{on} &&\Gamma \times (0,T),\\
  \rho\left(\boldsymbol{x},0\right) &= \rho_0\left(\boldsymbol{x}\right)\quad &&\text{in} &&\overline{\Omega},\\
  \boldsymbol{u}\left(\boldsymbol{x},0\right) &= \boldsymbol{u}_0\left(\boldsymbol{x}\right)\quad &&\text{in} &&\overline{\Omega},
\end{alignat*}
where $\boldsymbol{u}_0$ and $\rho_0$ are initial values of the
density and velocity, respectively. The function $\boldsymbol{f}$
represents the body force per unit mass. The viscous stress tensor is
given as
\[\boldsymbol{\tau}=\bar{\mu}\left(\nabla \boldsymbol{u} +\nabla^T \boldsymbol{u}\right) +\bar{\lambda} \nabla\cdot\boldsymbol{u}\boldsymbol{I}\]
where $\bar{\mu}$ and $\bar{\lambda}$ are the viscosity coefficients
and $\boldsymbol{I}$ is the identity tensor. The Korteweg tensor is
defined as
\[\boldsymbol{\zeta}=\lambda\left(\rho\Delta\rho+\frac{1}{2}\|\nabla\rho\|^2\right)\boldsymbol{I}-\lambda\nabla\rho\otimes\nabla\rho\]
where $\lambda$ is the capillarity coefficient. The pressure is given
by the van~der~Waals equation
\[p=Rb\frac{\rho\theta}{b-\rho}-a\rho^2,\] where $a,b$ are constants,
$R$ is the ideal gas constant, and $\theta$ is the temperature, which
for the isothermal model is assumed to be constant.

\begin{figure}
  \centering
  \subfloat[Density]{
    \begin{minipage}{0.98\textwidth}\centering
      \includegraphics[width=0.99\textwidth]{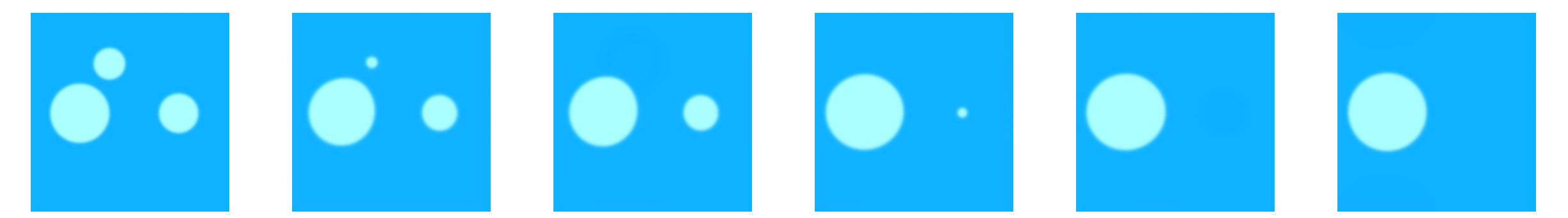}\\
      \includegraphics[width=0.4\textwidth]{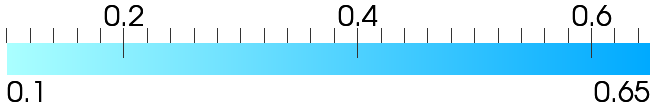}
    \end{minipage}
  }\\
  \subfloat[Velocity (magnitude)]{
    \begin{minipage}{0.98\textwidth}\centering
      \includegraphics[width=\textwidth]{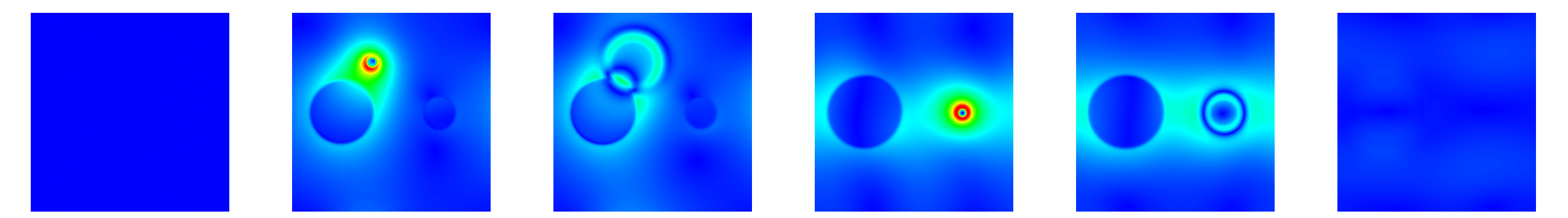}\\
      \includegraphics[width=0.4\textwidth]{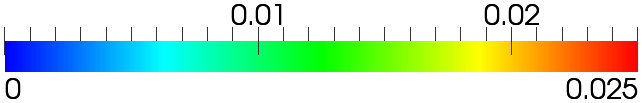}
    \end{minipage}}
  \caption{%
    Time evolution (left to right) of the density and the magnitude of
    the velocity for the isothermal NSK equations on the three bubble
    test case problem.}
  \label{fig:nsk2d}
\end{figure}

\begin{figure}
  \centering
  \subfloat[Near the beginning of the simulation]%
  {\includegraphics[width=0.46\textwidth]{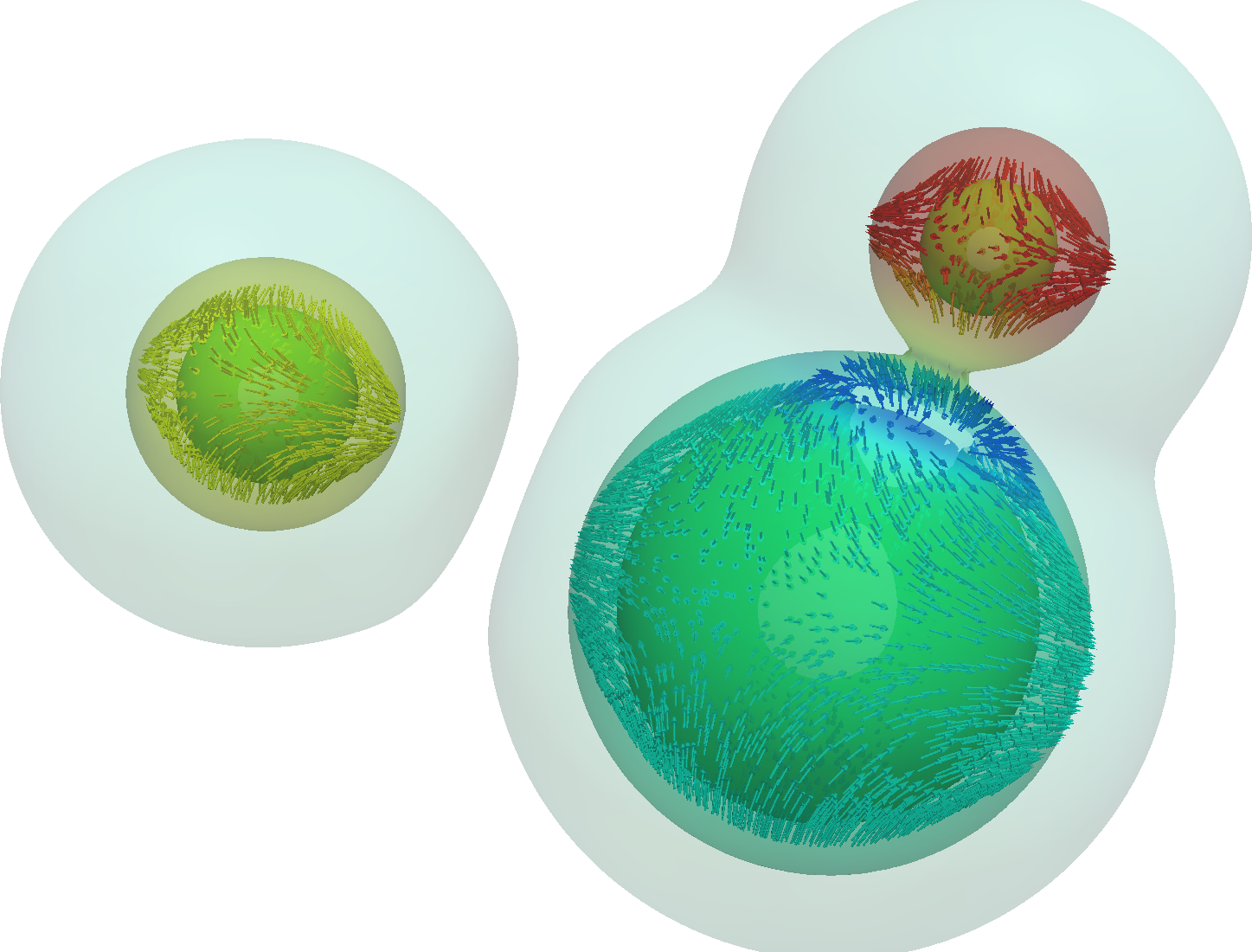}}%
  \hfill%
  \subfloat[Prior to the collapse of a bubble]%
  {\includegraphics[width=0.48\textwidth]{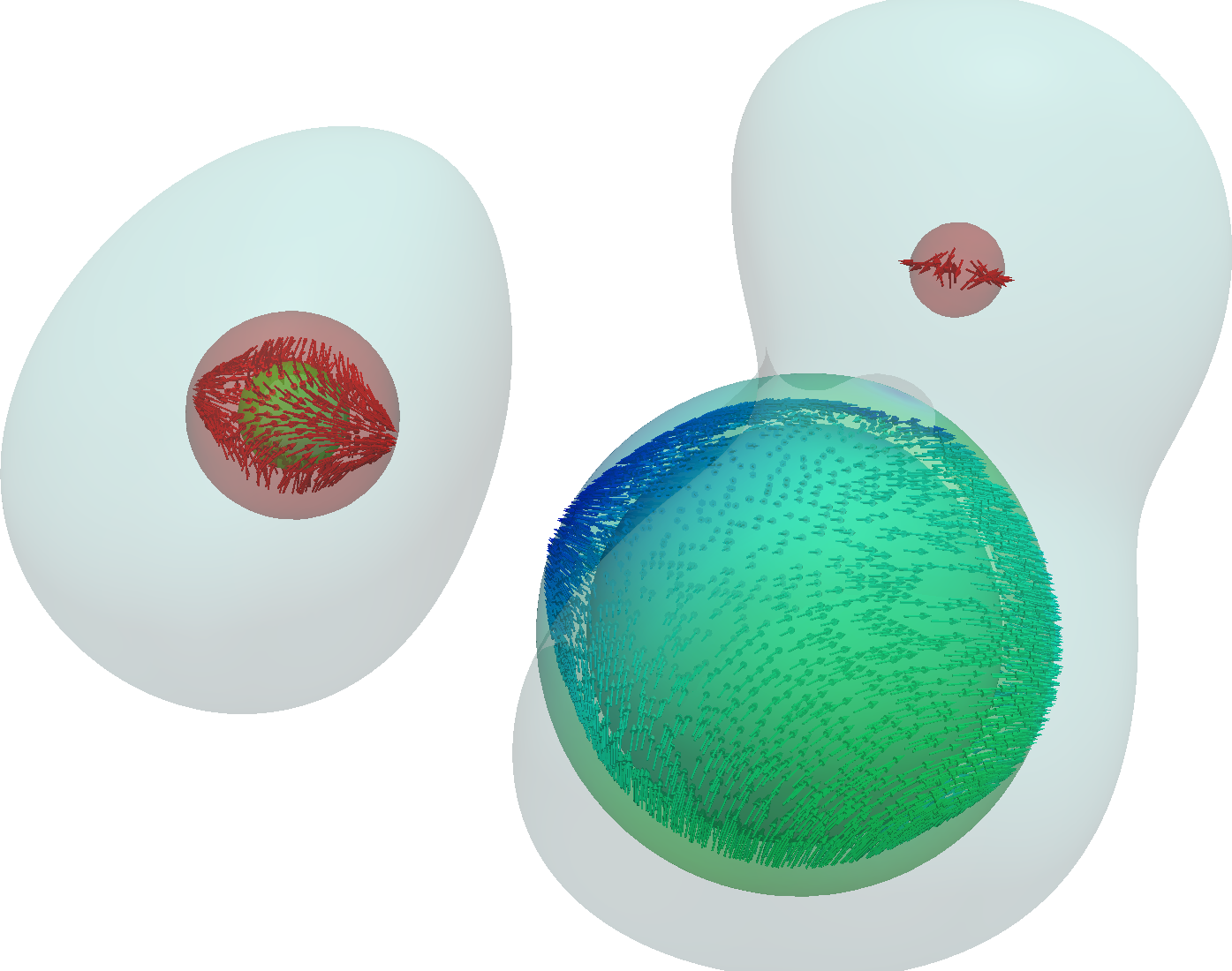}}%
  \caption{%
    Three-dimensional version of the three bubble problem for the
    Navier--Stokes--Korteweg equation. Isocontour surfaces reflect
    density values $\rho=\{0.15,0.55\}$ revealing the location of the
    three bubbles. Velocity vectors are shown on each isocontour and
    are colored by the velocity magnitude.}
  \label{fig:nsk3d}
\end{figure}

We solve the three-bubble test problem as detailed in \cite{Gomez2010}
and show here the density and magnitude of the velocity for a
two-dimensional solution in figure~\ref{fig:nsk2d} and a
three-dimensional solution in figure~\ref{fig:nsk3d}.

\subsubsection{Navier--Stokes Equations}

We solve the incompressible Navier--Stokes equations stabilized with
the variational multiscale method as formulated
in~\cite{Bazilevs2007}. Let $\Omega \in \mathbb{R}^d$ be an open set,
where $d=2,3$. The boundary of $\Omega$ with unit outward normal
$\boldsymbol{n}$ is denoted $\Gamma$.  The problem can be stated in
strong form as: find the velocity
$\boldsymbol{u}:\overline{\Omega}\times(0,T)\mapsto\mathbb{R}^d$ and
the pressure (divided by the constant density)
$p:\overline{\Omega}\times(0,T)\mapsto(0,b)$ such that
\begin{alignat*}{3}
  \dfrac{\partial\boldsymbol{u}}{\partial t} + \nabla\cdot\left(\boldsymbol{u}\otimes\boldsymbol{u}\right) + \nabla p &= \nu\Delta\boldsymbol{u} + \boldsymbol{f}\quad &&\text{in}\quad &&\Omega \times (0,T),\\
  \nabla\cdot\boldsymbol{u} &= 0\quad &&\text{in} &&\Omega \times (0,T),\\
  \boldsymbol{u} &= \boldsymbol{0}\quad &&\text{on} &&\Gamma \times (0,T),\\
  \boldsymbol{u}\left(\boldsymbol{x},0\right) &= \boldsymbol{u}_0\left(\boldsymbol{x}\right) &&\text{in} &&\overline{\Omega},
\end{alignat*}
where $\boldsymbol{u}_o$ is the initial velocity, $\boldsymbol{f}$
represents the body force per unit volume, and $\nu$ is the kinematic
viscosity.

\begin{figure}[h]
\centering
\subfloat[Perspective view]%
{\includegraphics[width=0.500\textwidth]{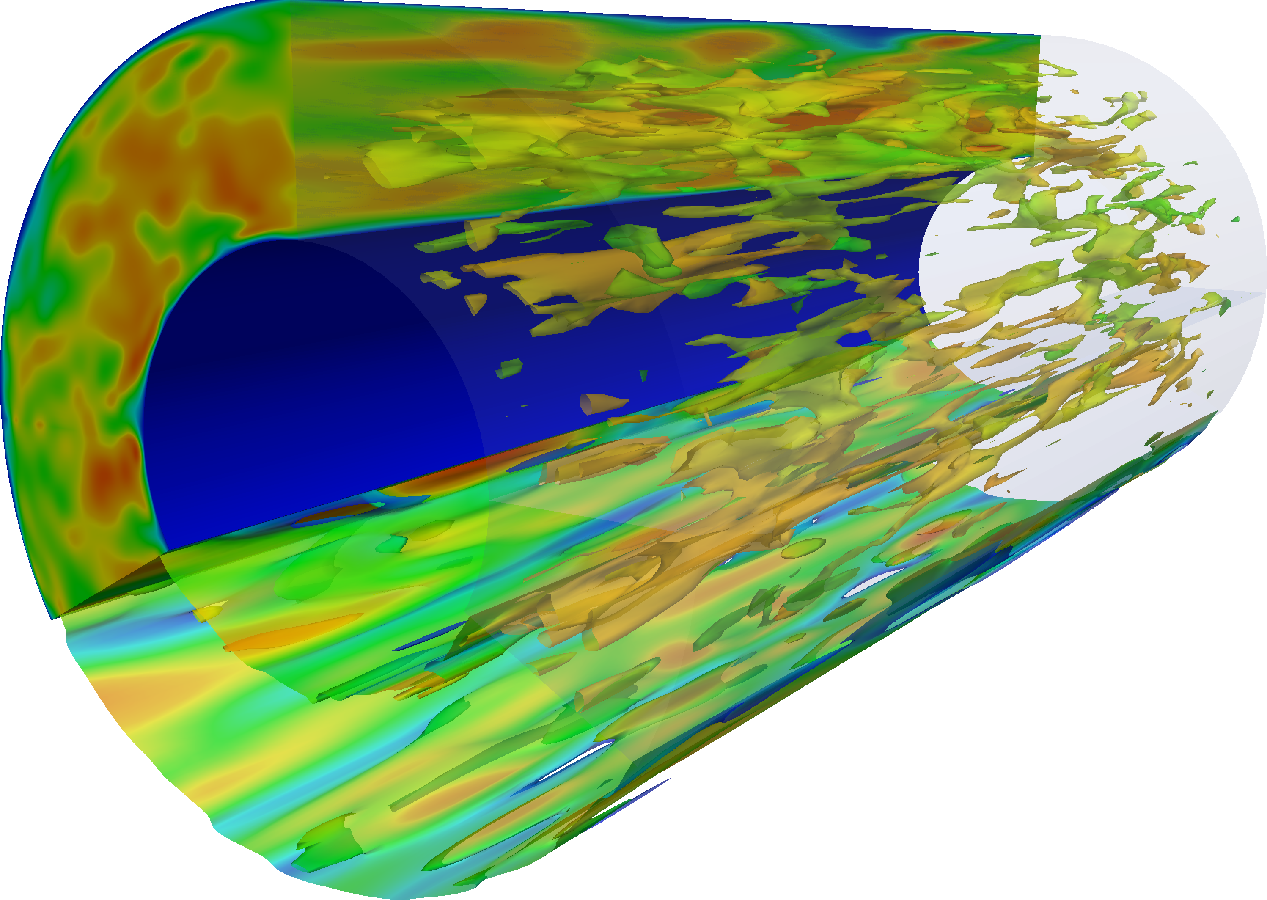}}%
\hfill%
\subfloat[Front view]%
{\includegraphics[width=0.375\textwidth]{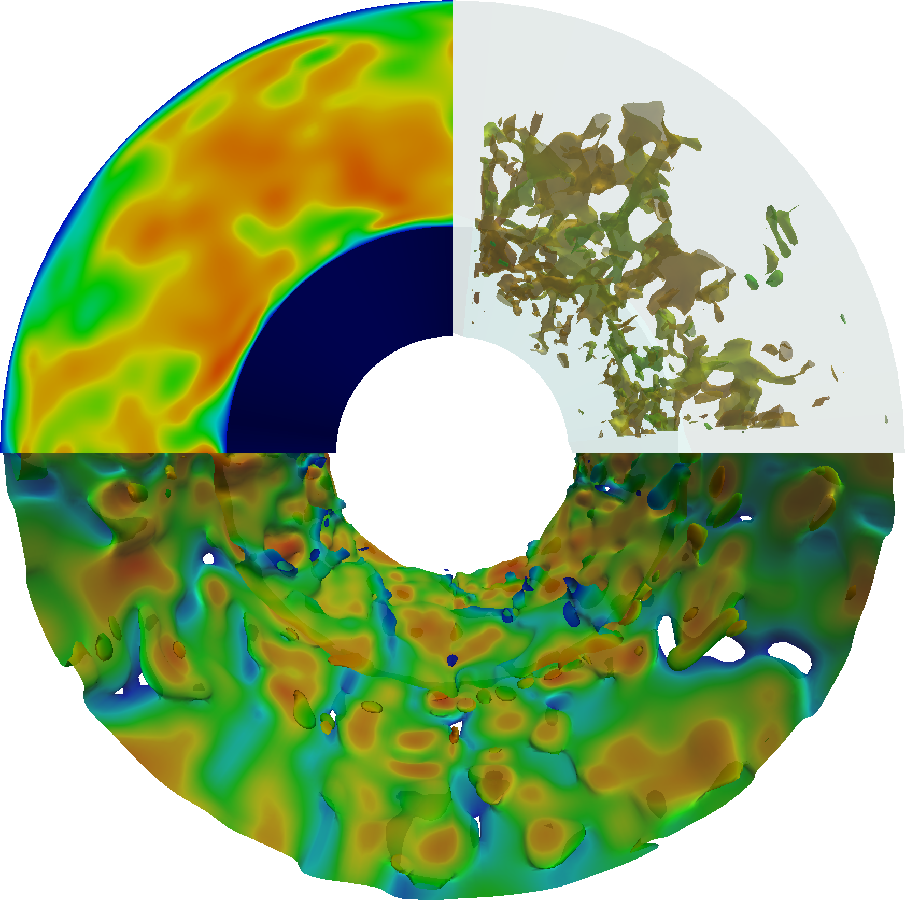}}%
\caption{%
  Turbulent flow through concentric cylinders as
  in~\cite{Motlagh2013}. The top-left quadrant is a pseudocolor plot
  of the streamwise velocity. The top-right quadrant shows isocontours
  of the vorticity magnitude for smaller values. The bottom quadrants
  shows isocontours of the vorticity magnitude for larger values. Both
  sets of contours are colored by the streamwise velocity.}
\label{fig:nsvms}
\end{figure}

We solve a turbulent flow in a concentric pipe as presented
in~\cite{Motlagh2013} the results of which we plot in
figure~\ref{fig:nsvms}.  The domain is periodic in the streamwise
direction. No-slip boundary conditions are set at the inner and outer
cylinder surfaces and the initial condition is set using a laminar
flow profile. The simulation is forced using a pressure gradient in
the form of a body force in the streamwise direction.

\section{Performance}\label{sec:performance}
In this section we present scaling results of our code framework,
applied to the incompressible Navier--Stokes equations described in
section~\ref{sec:applications}. Scaling of time-dependent, nonlinear
problems can be difficult to quantify as solver component performance
changes with problem size and decomposition. The nonlinearity
increases slightly with the problem size, which can lead to extra
Newton iterations for larger problems. The condition number of the
linear system increases considerably with the size of the problem,
requiring more linear iterations for convergence. At the same time,
domain-decomposition-based preconditioners become weaker as more
processors are employed, further increasing the number of iterations
required by the linear solver.

With the goal of measuring the performance of the code while
disregarding the algorithmic issues mentioned above, we perform tests
with the following assumptions and constraints
\begin{itemize}
\item The problem is the flow between two plates problem with no-slip
  boundary conditions at the plate walls.
\item The initial condition is set to the steady-state laminar flow
  profile.
\item We discretize the domain using $C^1$ quadratic B-splines and
  employ full Gauss--Legendre quadrature with three quadrature points
  per direction on each element.
\item We limit the test problem to ten time steps.
\item We enforce the use of two nonlinear Newton iterations per time step.
\item We enforce the use of $30$ GMRES iterations per Newton iteration.
\item The preconditioner is block Jacobi with one block per process.
  and ILU(0) (incomplete LU factorization with zero fill-in) in each
  block.
\end{itemize}

\begin{figure}[p]
\centering
\subfloat[Single node runs.]%
{\includegraphics[scale=0.6]{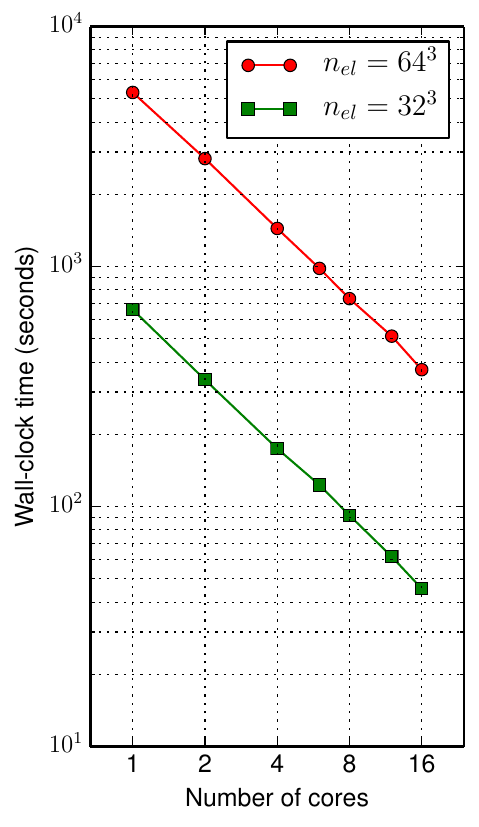}\label{fig:scaling-1}}%
\hfill%
\subfloat[Multiple node runs.]%
{\includegraphics[scale=0.6]{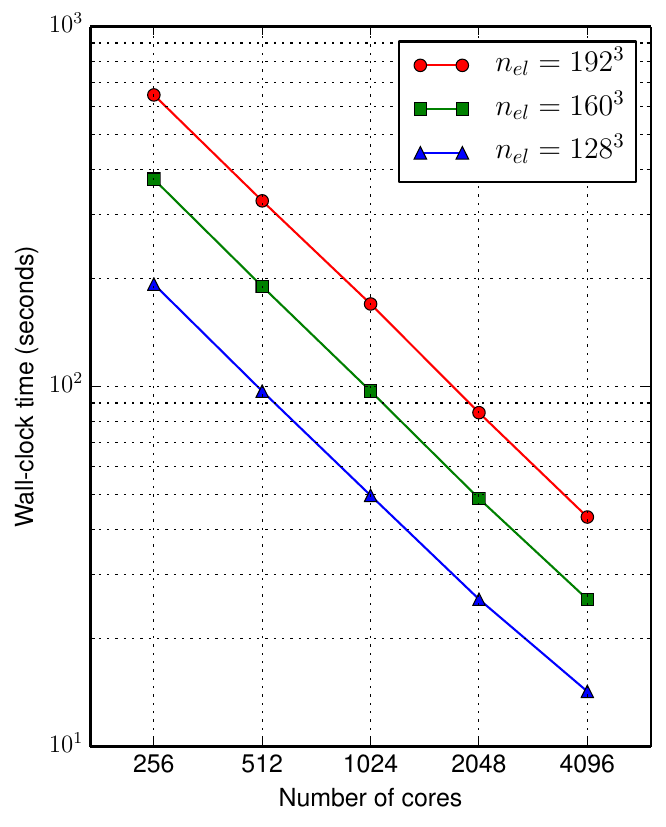}\label{fig:scaling-2}}%
\caption{%
  Strong scaling on Stampede. Wall-clock times corresponds to runs of
  the Navier--Stokes problem described in section~\ref{sec:applications}.
  Each run spans ten time steps with a fixed number of Newton
  and GMRES iterations.}
\label{fig:scaling}
\end{figure}

We ran these tests on Stampede~\cite{tacc}, a distributed memory
supercomputing system featuring compute nodes with 2 $\times$ 8-core
Intel Xeon E5-2680 (Sandy Bridge) processors, 32 GB of memory per
node, and a InfiniBand Mellanox interconnect. In
figure~\ref{fig:scaling-1} and table~\ref{tab:efficiency-1}, we show
strong scaling and efficiency of the Navier--Stokes code on a single
node using different discretization sizes indicated by the number of
elements used, $n_{el}$. In figure~\ref{fig:scaling-2} and
table~\ref{tab:efficiency-2}, we show strong scaling and efficiency on
multiple compute nodes. In these multiple node runs, the parallel
efficiency surpasses 80\%, and is above 90\% in all but one of the
cases. These values are achieved for local problem sizes
as small as $512$ elements per core.

\begin{table}[p]
\centering
\caption{Parallel efficiency on a single node of Stampede.}
\begin{tabular}{@{}cccccccc@{}}                                                   \toprule
               & \multicolumn{7}{c}{Number of cores}                           \\ \cmidrule(l){2-8}
 Mesh          & $1$     & $2$    & $4$    & $6$    & $8$    & $12$   & $16$   \\ \midrule
 $n_{el}=32^3$ & $100\%$ & $98\%$ & $95\%$ & $90\%$ & $90\%$ & $89\%$ & $91\%$ \\
 $n_{el}=64^3$ & $100\%$ & $94\%$ & $92\%$ & $90\%$ & $90\%$ & $86\%$ & $89\%$ \\ \bottomrule
\end{tabular}
\label{tab:efficiency-1}
\end{table}

\begin{table}[p]
\centering
\caption{Parallel efficiency on multiple nodes of Stampede.}
\begin{tabular}{@{}ccccccccc@{}}                                  \toprule
                & \multicolumn{5}{c}{Number of cores}          \\ \cmidrule(l){2-6}
 Mesh           & $256$   & $512$   & $1024$ & $2048$ & $4096$ \\ \midrule
 $n_{el}=128^3$ & $100\%$ & $99\%$  & $97\%$ & $94\%$ & $85\%$ \\
 $n_{el}=160^3$ & $100\%$ & $99\%$  & $97\%$ & $96\%$ & $92\%$ \\
 $n_{el}=192^3$ & $100\%$ & $98\%$  & $95\%$ & $95\%$ & $93\%$ \\ \bottomrule
\end{tabular}
\label{tab:efficiency-2}
\end{table}


\section{Conclusions}\label{sec:conclusions}
In this paper we present PetIGA, a scalable implementation of
isogeometric analysis for linear/nonlinear and static/transient
problems. By being built on top of PETSc, the framework gives users a
robust and versatile platform to solve partial differential
equations. We show that the framework scales well on up to $4096$
cores on the Navier--Stokes problem, and is therefore well suited for
large scale applications. Even though primarily conceived for
distributed-memory computing environments, PetIGA is also able to
extract excellent performance on nowadays shared memory multicore
laptop and desktop computers. Our software is already being used by
members of the community to tackle challenging problems related to
parallel multifrontal direct solvers~\cite{Wozniak2014}, fast
multipole-based preconditioners for elliptic
equations~\cite{Yokota2013}, multilevel Monte Carlo algorithms geared
towards the approximation of stochastic models~\cite{Collier2015},
fluid-structure interaction~\cite{Casquero2015}, and finite strain
gradient elasticity~\cite{Rudraraju2014}. 

Up to now, we focused on making single patch geometries run as
efficiently as possible. Single patch geometries are certainly enough
for a wide range of academic applications. However, support for
multipatch geometries is a must to tackle more complex
simulations. Still, extending our framework to support multiple
patches while keeping all of its user-friendly features and parallel
distributed memory capabilities is a nontrivial task that will be
addressed in future work. A multi-field extension of PetIGA is being
developed on top of the current implementation to handle discrete
differential forms. Promising results have already been obtained using
divergence- and integral-conforming spaces to solve the
Navier--Stokes--Cahn--Hilliard equation~\cite{Vignal2015}.


\section*{Acknowledgements}
We would like to acknowledge the open source software packages that
made this work possible: \texttt{PETSc}~\cite{petsc},
\texttt{NumPy}~\cite{numpy}, \texttt{matplotlib}~\cite{matplotlib},
\texttt{IPython}~\cite{ipython}. We would like to thank Lina Mar\'ia
Bernal Martinez, Gabriel Andres Espinosa Barrios, Federico
Fuentes-Caycedo, Juan Camilo Mahecha Zambrano for their work on the
hyper-elasticity implementation as a final project to the
\emph{Non-linear Finite Element Finite Element} class taught by
V.M.~Calo and N.~Collier for the Mechanical Engineering Department at
Universidad de Los Andes in Bogot\'{a}, Colombia in July 2012. We
would like to thank Adel Sarmiento Rodriguez for the visualization
work on figures~\ref{fig:nsk3d} and \ref{fig:nsvms}.

This work is part of the European Union's Horizon 2020 research and
innovation programme of the Marie Sk{\l}odowska-Curie grant agreement
No. 644602. This work was also supported by the Center for Numerical
Porous Media at King Abdullah University of Science and Technology.


\appendix
\pdfbookmark{Appendix}{appendix}
\section{Higher-order spatial derivatives}\label{app:ho-sp-ders}

The geometric mapping is computed using the
isoparametric concept, that is,
\begin{equation}
  \boldsymbol{x}(\boldsymbol{\xi})=\sum_A \boldsymbol{x}_A N_A(\boldsymbol{\xi}),
\end{equation}
where $N_A$ was defined in~\eqref{eq:rational} and $\boldsymbol{x}_A$
are the control point locations in physical space. Thus, the
parametric derivatives of this mapping are
\begin{align}
  \boldsymbol{x}_{,\alpha} &=
  \dfrac{\partial\boldsymbol{x}(\boldsymbol{\xi})}{\partial\xi_\alpha} =
  \sum_A \boldsymbol{x}_A N_{A,\alpha},\\
  \boldsymbol{x}_{,\alpha\beta} &=
  \dfrac{\partial\boldsymbol{x}(\boldsymbol{\xi})}{\partial\xi_\alpha\partial\xi_\beta} =
  \sum_A \boldsymbol{x}_A N_{A,\alpha\beta},\\
  \boldsymbol{x}_{,\alpha\beta\gamma} &=
  \dfrac{\partial\boldsymbol{x}(\boldsymbol{\xi})}{\partial\xi_\alpha\partial\xi_\beta\partial\xi_\gamma} =
  \sum_A \boldsymbol{x}_A N_{A,\alpha\beta\gamma}.
\end{align}
For simplicity, we write them in index notation as $x_{i,\alpha}$,
$x_{i,\alpha\beta}$, and $x_{i,\alpha\beta\gamma}$. Denoting the
inverse mapping as $\boldsymbol{\xi}(\boldsymbol{x})$, the first
spatial derivatives can be computed by matrix inversion through the
identity
\begin{align}
  \delta_{ij} = x_{i,\epsilon}\ \xi_{\epsilon,j},\label{eq:identity}
\end{align}
where $\delta_{ij}$ is the Kronecker delta. By repeated
differentiation, second and third spatial derivatives of the inverse
mapping follow:
\begin{align}
  \xi_{\alpha,ij} &= x_{m,\epsilon\mu}\ \xi_{\alpha,m}\ \xi_{\epsilon,i}\ \xi_{\mu,j}\label{eq:ddimap}\\
  \xi_{\alpha,ijk} &= -x_{m,\epsilon\mu\omega}\ \xi_{\alpha,m}\ \xi_{\epsilon,i}\ \xi_{\mu,j}\ \xi_{\omega,k}\notag\\
  &\quad-x_{m,\epsilon\mu}
  \left(
    \xi_{\alpha,mk} \ \xi_{\epsilon,i}  \ \xi_{\mu,j}  +
    \xi_{\alpha,m}  \ \xi_{\epsilon,ik} \ \xi_{\mu,j}  +
    \xi_{\alpha,m}  \ \xi_{\epsilon,i}  \ \xi_{\mu,jk}
  \right)\label{eq:dddimap}
\end{align}
Finally, the first, second, and third spatial derivatives of the basis
functions are
\begin{align}
  N_{A,i} &= N_{A,\alpha}\ \xi_{\alpha,i}\label{eq:dspatial}\\
  N_{A,ij} &= N_{A,\alpha\beta}\ \xi_{\alpha,i}\ \xi_{\beta,j} + N_{A,\alpha}\ \xi_{\alpha,ij}\label{eq:ddspatial}\\
  N_{A,ijk} &= N_{A,\alpha\beta\gamma}\ \xi_{\alpha,i}\ \xi_{\beta,j}\ \xi_{\gamma,k}\notag\\
  &+N_{A,\alpha\beta}\left(
    \xi_{\alpha,ik} \ \xi_{\beta,j}   +
    \xi_{\alpha,i}  \ \xi_{\beta,jk}  +
    \xi_{\alpha,ij} \ \xi_{\beta,k}
  \right)\notag\\
  &+N_{A,\alpha}\ \xi_{\alpha,ijk}.\label{eq:dddspatial}
\end{align}


\section{Tutorial for the Bratu example}\label{app:bratu-code}

\makeatletter
\let\verbatimdefaultfont\verbatim@font
\newcommand{\verbatimfont}[1]{\def\verbatim@font{#1}}%
\makeatother
\verbatimfont{\footnotesize\ttfamily}

This section complements section~\ref{sec:example}. It is meant as a
tutorial that focuses on building and running codes using PETSc and
PetIGA. In the following, we assume a POSIX environment such as
GNU/Linux or OS~X. Additionally, we assume that PETSc and PetIGA have
been properly configured and built with an MPI implementation to
enable parallel usage. As the usage of batch systems and job
schedulers is quite specific to the distributed-memory computing
environment users have access to, we only showcase parallel execution
within a single compute node or desktop/laptop computer.

We set environment variables pointing to the PETSc and PetIGA
directories as well as the PETSc build configuration. For recurrent
use, these variables can be defined in the shell configuration
file. When using the Bash shell, these definitions can be done as
follows
\begin{verbatim}
$ export PETSC_DIR=/path/to/petsc
$ export PETSC_ARCH=arch-platform-c-opt
$ export PETIGA_DIR=/path/to/petiga
\end{verbatim}

The \texttt{ls} command produces a listing with the content of the
current working directory
\begin{verbatim}
$ ls -lh
-rw-rw-r--. 1 user users 1.7K Dec 31 23:58 Bratu.c
-rw-rw-r--. 1 user users  242 Dec 31 23:58 makefile
\end{verbatim}
The \texttt{Bratu.c} file contains the C source code of the example
under consideration. This code is available at the end of this
section, see listing~\ref{lst:bratu}. The contents of
\texttt{makefile} can be inspected with the \texttt{cat} command
\begin{verbatim}
$ cat makefile
Bratu: Bratu.PETIGA;
include $(PETIGA_DIR)/lib/petiga/conf/variables
include $(PETIGA_DIR)/lib/petiga/conf/rules
\end{verbatim}
This \texttt{makefile} is simple: the first line defines a single
target and the following two lines include variable definitions and
build rules found within the PetIGA directory.

We use the \texttt{make} command to build the example. For the sake of
brevity, we omit the verbose output from the compiler and linker. We
then use the \texttt{ls} command once again to verify that the
\texttt{Bratu} binary executable has indeed been generated
\begin{verbatim}
$ make Bratu
...
$ ls -lh
-rwxrwxr-x. 1 user users 146K Dec 31 23:59 Bratu
-rw-rw-r--. 1 user users 1.7K Dec 31 23:58 Bratu.c
-rw-rw-r--. 1 user users  242 Dec 31 23:58 makefile
\end{verbatim}

Finally, we run the \texttt{Bratu} binary in parallel using four
processes using the \texttt{mpiexec} command. Additionally, we pass
some command line options to the program
\begin{verbatim}
$ mpiexec -n 4 ./Bratu -iga_elements 128 -iga_view -snes_monitor -ksp_type cg
IGA: dim=2 dof=1 order=2 geometry=0 rational=0 property=0
Axis 0: basis=BSPLINE[2,1] rule=LEGENDRE[3] periodic=0 nnp=130 nel=128
Axis 1: basis=BSPLINE[2,1] rule=LEGENDRE[3] periodic=0 nnp=130 nel=128
Partition - MPI: processors=[2,2,1] total=4
Partition - nnp: sum=16900 min=4096 max=4356 max/min=1.06348
Partition - nel: sum=16384 min=4096 max=4096 max/min=1
  0 SNES Function norm 5.266384548611e-02
  1 SNES Function norm 8.620220401724e-03
  2 SNES Function norm 2.054605014212e-03
  3 SNES Function norm 4.716226279209e-04
  4 SNES Function norm 8.916608064674e-05
  5 SNES Function norm 8.438014748365e-06
  6 SNES Function norm 1.155533195923e-07
  7 SNES Function norm 2.309601078808e-11
\end{verbatim}
The option \texttt{-iga\_view} displays various details related to the
discretization and parallel partitioning. From the first output line,
we verify the problem is two-dimensional and the solution has a single
component. The second and third line provide information about the
discretization along each direction: the problem is being solved in a
grid with $128{\times}128$ elements (as requested through the %
\texttt{-iga\_elements} option), quadratic $C^1$ B-spline basis %
functions, and Gauss--Legendre quadrature with three points per %
direction.  The option \texttt{-snes\_monitor} monitors the nonlinear
solver convergence by printing the iteration number and the 2-norm of
the global residual vector. Finally, the option \texttt{-ksp\_type}
configures the linear solver to use the conjugate gradient
method. Many more options are available to control the nonlinear
solver, the linear solver, and the preconditioner. These options can
be listed by passing the \texttt{-help} option.

\begin{figure}
  \centering
  \subfloat[Visualizing at runtime.]      {\includegraphics[width=0.35\textwidth]{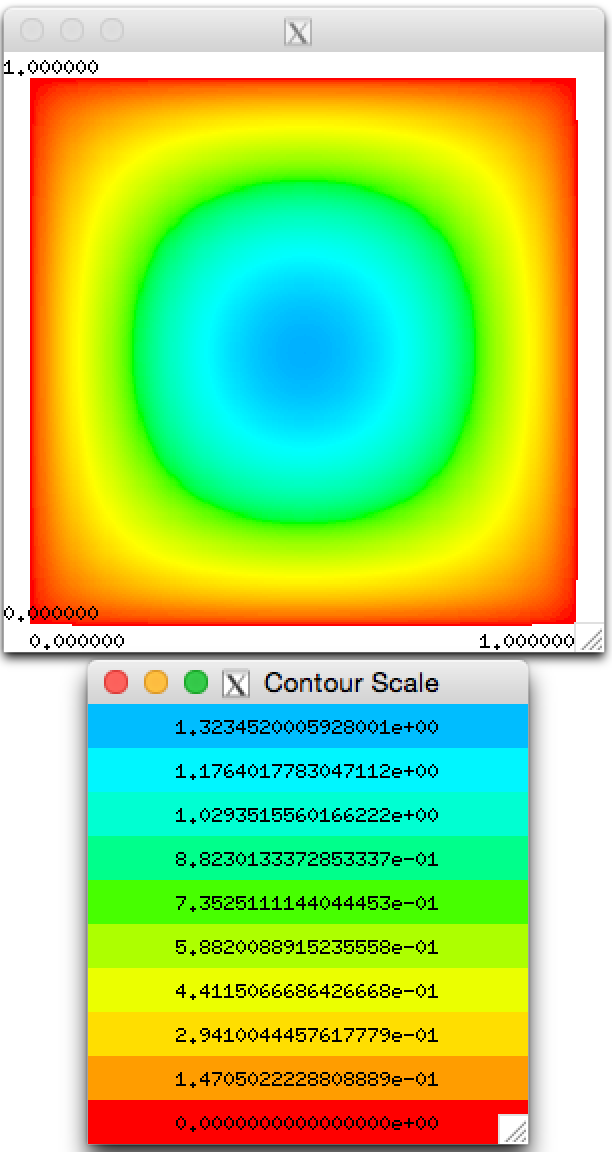}\label{fig:bratuxeleven}}~
  \subfloat[Post-processing with ParaView.]{\includegraphics[width=0.62\textwidth]{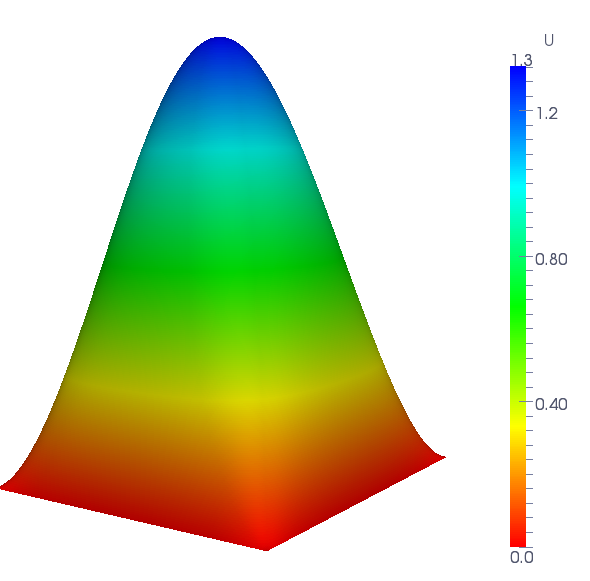}\label{fig:bratuparaview}}\\
  \caption{%
    Visualizing the solution of the Bratu equation.}
  \label{fig:Bratu}
\end{figure}

Thanks to the call to the \texttt{VecViewFromOptions} routine, we
visualize the solution at runtime %
\begin{verbatim}
$ mpiexec -n 4 ./Bratu -iga_elements 128 -output draw:x -draw_pause 5
\end{verbatim}
This feature allows users to quickly check for a sensible solution, as
seen in figure~\ref{fig:bratuxeleven}. This is particularly useful
while in the initial setup of a time-dependent problem. To obtain
higher quality visualizations, the option value can be changed to dump
a VTK file to disk, which can then be post-processed with tools such
as ParaView~\cite{paraview}, as seen in
figure~\ref{fig:bratuparaview}.
\begin{verbatim}
$ mpiexec -n 4 ./Bratu -iga_elements 128 -output vtk:Bratu.vts
$ ls -lh Bratu.vts
-rw-rw-r--. 1 user users 2.1M Jan  1 00:01 Bratu.vts
$ paraview Bratu.vts
\end{verbatim}

\codelisting[
float=p,%
caption={C source code for the Bratu example},%
label=lst:bratu,
basicstyle=\scriptsize\ttfamily,
numbers=none,
frame=none,
lineskip={-1.5pt},%
]{1}{72}{lst-Bratu.c}


\section{Numerical differentiation comparison}\label{app:num-diff}

In this section we present some benchmark results measuring the
computational overhead of two numerical differentiation approaches
discussed in section~\ref{sec:num-diff}. We base our tests on the
Bratu problem described in section~\ref{sec:example}. However, we now
focus on the three-dimensional counterpart to make the problem more
computationally challenging. We use a modified version of the code
listing~\ref{lst:bratu} presented in appendix~\ref{app:bratu-code}:
instead of solving the nonlinear problem, we compute the global
Jacobian matrix and report the minimum wall clock time from five
successive computations. These computations are performed with three
different methods.
\begin{itemize}
\item \emph{Explicit} uses the explicitly coded, user-provided
  Jacobian routine, as described in section~\ref{sec:example},
\item \emph{Coloring} uses the global colored finite differences
  implementation available in PETSc,
\item \emph{Local~ND} uses the local (i.e., at quadrature points)
  numerical differentiation implementation available in PetIGA.
\end{itemize}
The benchmarks where run in a workstation with two Intel Xeon CPU
E5-2680 v2 (10 cores, 2.80GHz, 25 MB cache, 8 GT/s QPI) and 128 GB of
memory running Linux~4.0.4 (Fedora~21) and used the in-development
versions of PETSc and PetIGA compiled with GCC~4.9.2 using the
\texttt{-Ofast} optimization flag.

The benchmark results are summarized in table~\ref{tab:num-diff}.
Parameters $P$, $n_{el}$, $p$, and $k$ refer to number of processes,
number of elements, degree and continuity order of the polynomial
space, respectively. All cases used full Gauss--Legendre quadrature
with $p+1$ quadrature points per direction. Taking the explicitly
coded Jacobian computations as a baseline, colored finite differences
is over an order of magnitude slower whereas the local numerical
differentiation approach is only about four times slower.

\begin{table}[h]
\centering
\caption{%
  Benchmark of numerical differentiation schemes. Parameters $P$ and $n_{el}$
  denote number of processes and elements, respectively, while $p$
  and $k$ denote polynomial degree and continuity order, respectively.
  We report the minimum wall clock time from five successive computations
  of the global Jacobian matrix corresponding to the 3D Bratu problem.
}\label{tab:num-diff}
\newcolumntype{.}{D{.}{.}{-1}}
\newcommand{\mrw}[1]{\multirow{2}{*}[-0.5ex]{#1}}
\newcommand{\col}[1]{\multicolumn{1}{c}{\text{#1}}}
\begin{tabular}{@{}cccc...@{}}                                                            \toprule
    &               &         &     & \multicolumn{3}{c}{Time (seconds)}               \\ \cmidrule[\heavyrulewidth](l){5-7}
$P$ & $n_{el}$      & $p$     & $k$ & \col{Explicit} & \col{Coloring} & \col{Local~ND} \\ \midrule
    &               & \mrw{2} & 0   &   0.31         &   7.67         &  1.21          \\ \cmidrule(l){4-7}
8   & $32^3$        &         & 1   &   0.33         &   7.79         &  1.24          \\ \cmidrule(l){3-7}
    &               & 3       & 2   &   3.71         & 105.08         & 15.27          \\ \midrule
    &               & \mrw{2} & 0   &   1.31         &  34.44         &  5.15          \\ \cmidrule(l){4-7}
    & $64^3$        &         & 1   &   1.35         &  34.84         &  5.26          \\ \cmidrule(l){3-7}
16  &               & 3       & 2   &  15.41         & 452.67         & 64.93          \\ \cmidrule(l){2-7}
    & \mrw{$128^3$} & 1       & 0   &   0.52         &  10.35         &  1.72          \\ \cmidrule(l){3-7}
    &               & 2       & 1   &  10.74         & 307.06         & 41.46          \\ \bottomrule
\end{tabular}
\end{table}


\clearpage
\pdfbookmark{References}{references}
\biboptions{numbers,sort&compress}
\bibliographystyle{paper}
\bibliography{paper}

\begin{thebibliography}{10}
\expandafter\ifx\csname url\endcsname\relax
  \def\url#1{\texttt{#1}}\fi
\expandafter\ifx\csname urlprefix\endcsname\relax\def\urlprefix{URL }\fi
\expandafter\ifx\csname href\endcsname\relax
  \def\href#1#2{#2} \def\path#1{#1}\fi

\bibitem{Hughes2005}
T.J.R. Hughes, J.A. Cottrell, Y.~Bazilevs, Isogeometric analysis: {CAD}, finite
  elements, {NURBS}, exact geometry and mesh refinement, Computer Methods in
  Applied Mechanics and Engineering 194~(39--41) (2005) 4135--4195.
\newblock \href {http://dx.doi.org/10.1016/j.cma.2004.10.008}
  {\path{doi:10.1016/j.cma.2004.10.008}}.

\bibitem{Cottrell2009}
J.A. Cottrell, T.J.R. Hughes, Y.~Bazilevs, Isogeometric Analysis: Toward
  Unification of {CAD} and {FEA}, John Wiley \& Sons, Ltd, 2009.
\newblock \href {http://dx.doi.org/10.1002/9780470749081}
  {\path{doi:10.1002/9780470749081}}.

\bibitem{Bercovier1987}
M.~Bercovier, G.~Berold, Solving design problems by integration of {CAD} and
  {FEM} software, in: B.~Ford, F.~Chatelin (Eds.), Proceedings of IFIP TC 2/WG
  2.5 Working Conference on Problem Solving Environments for Scientific
  Computing, Sophia Antipolis, France, 17--21 June, 1985, North Holland, 1987,
  p. 309.

\bibitem{Sheffer1998}
A.~Sheffer, T.~Blacker, M.~Bercovier, Steps towards smooth {CAD--FEM}
  integration, in: Proceedings of 6th International Conference on Numerical
  Grid Generation in Computational Field Simulations, 1998, pp. 705--714.

\bibitem{Sheffer1999}
A.~Sheffer, M.~Bercovier, {CAD} model editing and its applications, Ph.D.
  thesis, Hebrew University of Jerusalem (1999).

\bibitem{Hollig2003}
K.~H\"ollig, Finite Element Methods with {B}-splines, Frontiers in Applied
  Mathematics, Society for Industrial and Applied Mathematics, 2003.
\newblock \href {http://dx.doi.org/10.1137/1.9780898717532}
  {\path{doi:10.1137/1.9780898717532}}.

\bibitem{Gomez2008}
H.~Gomez, V.M. Calo, Y.~Bazilevs, T.J.R. Hughes, Isogeometric analysis of the
  {Cahn--Hilliard} phase-field model, Computer Methods in Applied Mechanics and
  Engineering 197~(49--50) (2008) 4333--4352.
\newblock \href {http://dx.doi.org/10.1016/j.cma.2008.05.003}
  {\path{doi:10.1016/j.cma.2008.05.003}}.

\bibitem{Gomez2010}
H.~Gomez, T.J.R. Hughes, X.~Nogueira, V.M. Calo, Isogeometric analysis of the
  isothermal {Navier--Stokes--Korteweg} equations, Computer Methods in Applied
  Mechanics and Engineering 199~(25--28) (2010) 1828--1840.
\newblock \href {http://dx.doi.org/10.1016/j.cma.2010.02.010}
  {\path{doi:10.1016/j.cma.2010.02.010}}.

\bibitem{Vignal2015b}
P.~Vignal, L.~Dalcin, D.L. Brown, N.~Collier, V.M. Calo, An energy-stable
  convex splitting for the phase-field crystal equation, Computers \&
  Structures 158 (2015) 355–--368.
\newblock \href {http://dx.doi.org/10.1016/j.compstruc.2015.05.029}
  {\path{doi:10.1016/j.compstruc.2015.05.029}}.

\bibitem{Benson2011}
D.J. Benson, Y.~Bazilevs, M.-C. Hsu, T.J.R. Hughes, A large deformation,
  rotation-free, isogeometric shell, Computer Methods in Applied Mechanics and
  Engineering 200~(13--16) (2011) 1367--1378.
\newblock \href {http://dx.doi.org/10.1016/j.cma.2010.12.003}
  {\path{doi:10.1016/j.cma.2010.12.003}}.

\bibitem{Kiendl2015}
J.~Kiendl, M.-C. Hsu, M.C.H. Wu, A.~Reali, Isogeometric {K}irchhoff--{L}ove
  shell formulations for general hyperelastic materials, Computer Methods in
  Applied Mechanics and Engineering 291~(0) (2015) 280--303.
\newblock \href {http://dx.doi.org/10.1016/j.cma.2015.03.010}
  {\path{doi:10.1016/j.cma.2015.03.010}}.

\bibitem{Bouclier2015}
R.~Bouclier, T.~Elguedj, A.~Combescure, An isogeometric locking--free
  {NURBS}-based solid--shell element for geometrically nonlinear analysis,
  International Journal for Numerical Methods in Engineering 101~(10) (2015)
  774--808.
\newblock \href {http://dx.doi.org/10.1002/nme.4834}
  {\path{doi:10.1002/nme.4834}}.

\bibitem{Akkerman2008}
I.~Akkerman, Y.~Bazilevs, V.M. Calo, T.J.R. Hughes, S.~Hulshoff, The role of
  continuity in residual-based variational multiscale modeling of turbulence,
  Computational Mechanics 41~(3) (2008) 371--378.
\newblock \href {http://dx.doi.org/10.1007/s00466-007-0193-7}
  {\path{doi:10.1007/s00466-007-0193-7}}.

\bibitem{Akkerman2011}
I.~Akkerman, Y.~Bazilevs, C.E. Kees, M.W. Farthing, Isogeometric analysis of
  free-surface flow, Journal of Computational Physics 230~(11) (2011)
  4137--4152, special issue High Order Methods for CFD Problems.
\newblock \href {http://dx.doi.org/10.1016/j.jcp.2010.11.044}
  {\path{doi:10.1016/j.jcp.2010.11.044}}.

\bibitem{Buffa2011}
A.~Buffa, J.~Rivas, G.~Sangalli, R.~V{\'{a}}zquez, Isogeometric discrete
  differential forms in three dimensions, SIAM Journal on Numerical Analysis
  49~(2) (2011) 818--844.
\newblock \href {http://dx.doi.org/10.1137/100786708}
  {\path{doi:10.1137/100786708}}.

\bibitem{Evans2013}
J.A. Evans, T.J.R. Hughes, Isogeometric divergence-conforming {B}-splines for
  the unsteady {N}avier--{S}tokes equations, Journal of Computational Physics
  241 (2013) 141--167.
\newblock \href {http://dx.doi.org/10.1016/j.jcp.2013.01.006}
  {\path{doi:10.1016/j.jcp.2013.01.006}}.

\bibitem{Veiga2011}
L.~Beir{\~a}o da~Veiga, A.~Buffa, J.~Rivas, G.~Sangalli, Some estimates for
  h--p--k-refinement in {I}sogeometric {A}nalysis, Numerische Mathematik 118
  (2011) 271--305.
\newblock \href {http://dx.doi.org/10.1007/s00211-010-0338-z}
  {\path{doi:10.1007/s00211-010-0338-z}}.

\bibitem{Evans2009}
J.A. Evans, Y.~Bazilevs, I.~Babu{\v{s}}ka, T.J.R. Hughes, $n$-{W}idths,
  sup--infs, and optimality ratios for the $k$-version of the isogeometric
  finite element method, Computer Methods in Applied Mechanics and Engineering
  198~(21--26) (2009) 1726--1741, advances in Simulation-Based Engineering
  Sciences -- Honoring J. Tinsley Oden.
\newblock \href {http://dx.doi.org/10.1016/j.cma.2009.01.021}
  {\path{doi:10.1016/j.cma.2009.01.021}}.

\bibitem{Collier2014}
N.~Collier, L.~Dalcin, V.M. Calo, On the computational efficiency of
  isogeometric methods for smooth elliptic problems using direct solvers,
  International Journal for Numerical Methods in Engineering 100~(8) (2014)
  620--632.
\newblock \href {http://dx.doi.org/10.1002/nme.4769}
  {\path{doi:10.1002/nme.4769}}.

\bibitem{Collier2012}
N.~Collier, D.~Pardo, L.~Dalcin, M.~Paszynski, V.M. Calo, The cost of
  continuity: A study of the performance of isogeometric finite elements using
  direct solvers, Computer Methods in Applied Mechanics and Engineering
  213--216 (2012) 353--361.
\newblock \href {http://dx.doi.org/10.1016/j.cma.2011.11.002}
  {\path{doi:10.1016/j.cma.2011.11.002}}.

\bibitem{Collier2013}
N.~Collier, L.~Dalcin, D.~Pardo, V.M. Calo, The cost of continuity: Performance
  of iterative solvers on isogeometric finite elements, SIAM Journal on
  Scientific Computing 35~(2) (2013) A767--A784.
\newblock \href {http://dx.doi.org/10.1137/120881038}
  {\path{doi:10.1137/120881038}}.

\bibitem{Baxter2012}
R.~Baxter, N.C. Hong, D.~Gorissen, J.~Hetherington, I.~Todorov, The research
  software engineer, in: Digital Research Conference, Oxford, 2012, pp. 1--3.

\bibitem{Wilson2008}
G.~Wilson, Those who will not learn from history..., Computing in Science \&
  Engineering 10~(3) (2008) 5--6.
\newblock \href {http://dx.doi.org/10.1109/MCSE.2008.86}
  {\path{doi:10.1109/MCSE.2008.86}}.

\bibitem{petsc}
S.~Balay, S.~Abhyankar, M.F. Adams, J.~Brown, P.~Brune, K.~Buschelman,
  L.~Dalcin, V.~Eijkhout, W.D. Gropp, D.~Kaushik, M.G. Knepley, L.~Curfman
  McInnes, K.~Rupp, B.F. Smith, S.~Zampini, H.~Zhang.
\newblock \href{http://www.mcs.anl.gov/petsc}{{PETS}c {W}eb page} [online]
  (2015).
\newline\urlprefix\url{http://www.mcs.anl.gov/petsc}

\bibitem{petsc-manual}
S.~Balay, S.~Abhyankar, M.F. Adams, J.~Brown, P.~Brune, K.~Buschelman,
  L.~Dalcin, V.~Eijkhout, W.D. Gropp, D.~Kaushik, M.G. Knepley, L.~Curfman
  McInnes, K.~Rupp, B.F. Smith, S.~Zampini, H.~Zhang,
  \href{http://www.mcs.anl.gov/petsc/petsc-current/docs/manual.pdf}{{PETSc}
  users manual}, Tech. Rep. ANL-95/11 - Revision 3.6, Argonne National
  Laboratory (2015).
\newline\urlprefix\url{http://www.mcs.anl.gov/petsc/petsc-current/docs/manual.pdf}

\bibitem{petsc-efficient}
S.~Balay, W.D. Gropp, L.~Curfman McInnes, B.F. Smith, Efficient management of
  parallelism in object oriented numerical software libraries, in: E.~Arge,
  A.M. Bruaset, H.P. Langtangen (Eds.), Modern Software Tools in Scientific
  Computing, Birkh{\"{a}}user Press, 1997, pp. 163--202.

\bibitem{hypre}
R.D. Falgout, U.M. Yang, \textit{hypre}: A library of high performance
  preconditioners, in: P.M.A. Sloot, A.G. Hoekstra, C.J.~Kenneth Tan, J.J.
  Dongarra (Eds.), Computational Science -- ICCS 2002, Vol. 2331 of Lecture
  Notes in Computer Science, Springer Berlin Heidelberg, 2002, pp. 632--641.
\newblock \href {http://dx.doi.org/10.1007/3-540-47789-6_66}
  {\path{doi:10.1007/3-540-47789-6_66}}.

\bibitem{trilinos}
M.A. Heroux, R.A. Bartlett, V.E. Howle, R.J. Hoekstra, J.J. Hu, T.G. Kolda,
  R.B. Lehoucq, K.R. Long, R.P. Pawlowski, E.T. Phipps, A.G. Salinger, H.K.
  Thornquist, R.S. Tuminaro, J.M. Willenbring, A.~Williams, K.S. Stanley, An
  overview of the {T}rilinos project, ACM Transactions on Mathematical Software
  31~(3) (2005) 397--423.
\newblock \href {http://dx.doi.org/10.1145/1089014.1089021}
  {\path{doi:10.1145/1089014.1089021}}.

\bibitem{mumps}
P.R. Amestoy, A.~Guermouche, J.-Y. L'Excellent, S.~Pralet, Hybrid scheduling
  for the parallel solution of linear systems, Parallel Computing 32~(2) (2006)
  136--156, parallel Matrix Algorithms and Applications.
\newblock \href {http://dx.doi.org/10.1016/j.parco.2005.07.004}
  {\path{doi:10.1016/j.parco.2005.07.004}}.

\bibitem{dealII}
W.~Bangerth, R.~Hartmann, G.~Kanschat, {deal.II} -- a general purpose object
  oriented finite element library, ACM Transactions on Mathematical Software
  33~(4) (2007) 24/1--24/27.
\newblock \href {http://dx.doi.org/10.1145/1268776.1268779}
  {\path{doi:10.1145/1268776.1268779}}.

\bibitem{fenics}
A.~Logg, K.-A. Mardal, G.~Wells, Automated Solution of Differential Equations
  by the Finite Element Method, Springer Berlin Heidelberg, 2012.

\bibitem{libmesh}
B.S. Kirk, J.W. Peterson, R.H. Stogner, G.F. Carey, {\texttt{libMesh}: {A}
  {C++} Library for Parallel Adaptive Mesh Refinement/Coarsening Simulations},
  Engineering with Computers 22~(3--4) (2006) 237--254.
\newblock \href {http://dx.doi.org/10.1007/s00366-006-0049-3}
  {\path{doi:10.1007/s00366-006-0049-3}}.

\bibitem{petscfem}
V.E. Sonzogni, A.M. Yommi, N.M. Nigro, M.A. Storti, A parallel finite element
  program on a beowulf cluster, Advances in Engineering Software 33~(7--10)
  (2002) 427--443, engineering Computational Technology \& Computational
  Structures Technology.
\newblock \href {http://dx.doi.org/10.1016/S0965-9978(02)00059-5}
  {\path{doi:10.1016/S0965-9978(02)00059-5}}.

\bibitem{petiga}
L.~Dalcin, N.~Collier.
\newblock \href{https://bitbucket.org/dalcinl/petiga}{{PetIGA}: High
  performance isogeometric analysis} [online] (2015).
\newline\urlprefix\url{https://bitbucket.org/dalcinl/petiga}

\bibitem{Cox1972}
M.G. Cox, The numerical evaluation of {B}-splines, IMA Journal of Applied
  Mathematics 10~(2) (1972) 134--149.
\newblock \href {http://dx.doi.org/10.1093/imamat/10.2.134}
  {\path{doi:10.1093/imamat/10.2.134}}.

\bibitem{DeBoor1972}
C.~de~Boor, On calculation with {B}-splines, Journal of Approximation Theory
  6~(1) (1972) 50--62.
\newblock \href {http://dx.doi.org/10.1016/0021-9045(72)90080-9}
  {\path{doi:10.1016/0021-9045(72)90080-9}}.

\bibitem{Piegl1995}
L.~Piegl, W.~Tiller, The {NURBS} Book, Monographs in Visual Communication,
  Springer, 1995.

\bibitem{Hughes2000}
T.J.R. Hughes, The finite element method: linear static and dynamic finite
  element analysis, Dover Publications, 2000.

\bibitem{Liu2013}
J.~Liu, L.~Ded\`{e}, J.A. Evans, M.J. Borden, T.J.R. Hughes, Isogeometric
  analysis of the advective {C}ahn-{H}illiard equation: Spinodal decomposition
  under shear flow, Journal of Computational Physics 242 (2013) 321--350.
\newblock \href {http://dx.doi.org/10.1016/j.jcp.2013.02.008}
  {\path{doi:10.1016/j.jcp.2013.02.008}}.

\bibitem{Knoll2004}
D.A. Knoll, D.E. Keyes, Jacobian-free newton–krylov methods: a survey of
  approaches and applications, Journal of Computational Physics 193~(2) (2004)
  357--397.
\newblock \href {http://dx.doi.org/10.1016/j.jcp.2003.08.010}
  {\path{doi:10.1016/j.jcp.2003.08.010}}.

\bibitem{Gebremedhin2005}
A.H. Gebremedhin, F.~Manne, A.~Pothen, What color is your {J}acobian? graph
  coloring for computing derivatives, SIAM Review 47~(4) (2005) 629--705.
\newblock \href {http://dx.doi.org/10.1137/S0036144504444711}
  {\path{doi:10.1137/S0036144504444711}}.

\bibitem{Pernice1998}
M.~Pernice, H.~F. Walker, {NITSOL}: A {N}ewton iterative solver for nonlinear
  systems, SIAM Journal on Scientific Computing 19 (1998) 302--318.

\bibitem{Hughes2010}
T.J.R. Hughes, A.~Reali, G.~Sangalli, Efficient quadrature for {NURBS}-based
  isogeometric analysis, Computer Methods in Applied Mechanics and Engineering
  199~(5--8) (2010) 301--313.

\bibitem{AitHaddou2014}
R.~Ait-Haddou, M.~Barto{\v{n}}, V.M. Calo, Explicit {G}aussian quadrature rules
  for cubic splines with non-uniform knot sequences, ArXiv e-prints (2014)
  1--15.\href {http://arxiv.org/abs/1410.7196} {\path{arXiv:1410.7196}}.

\bibitem{Barton2015a}
M.~Barto{\v{n}}, R.~Ait-Haddou, V.M. Calo, {G}aussian quadrature rules for
  {$C^1$} quintic splines, ArXiv e-prints (2015) 1--20.\href
  {http://arxiv.org/abs/1503.00907} {\path{arXiv:1503.00907}}.

\bibitem{Barton2015b}
M.~Barto{\v{n}}, V.M. Calo, {G}aussian quadrature for splines via homotopy
  continuation: rules for {$C^2$} cubic splines, ArXiv e-prints (2015)
  1--22.\href {http://arxiv.org/abs/1505.04391} {\path{arXiv:1505.04391}}.

\bibitem{Karniadakis2013}
G.E. Karniadakis, S.J. Sherwin, Spectral/hp Element Methods for Computational
  Fluid Dynamics, 2nd Edition, Oxford University Press, 2013.

\bibitem{Auricchio2010}
F.~Auricchio, Beir{\~{a}}o~Da Veiga, T.J.R. Hughes, A.~Reali, G.~Sangalli,
  Isogeometric collocation methods, Mathematical Models and Methods in Applied
  Sciences 20~(11) (2010) 2075--2107.
\newblock \href {http://dx.doi.org/10.1142/S0218202510004878}
  {\path{doi:10.1142/S0218202510004878}}.

\bibitem{igakit}
L.~Dalcin, N.~Collier.
\newblock \href{https://bitbucket.org/dalcinl/igakit}{igakit} [online] (2015).
\newline\urlprefix\url{https://bitbucket.org/dalcinl/igakit}

\bibitem{igakit-tutorial}
N.~Collier, L.~Dalcin.
\newblock \href{https://petiga-igakit.readthedocs.org}{{PetIGA} and {igakit}
  tutorial} [online] (2015).
\newline\urlprefix\url{https://petiga-igakit.readthedocs.org}

\bibitem{Jacobsen2002}
J.~Jacobsen, K.~Schmitt, The {L}iouville-{B}ratu-{G}elfand {P}roblem for
  {R}adial {O}perators, Journal of {D}ifferential {E}quations 184~(1) (2002)
  283--298.
\newblock \href {http://dx.doi.org/10.1006/jdeq.2001.4151}
  {\path{doi:10.1006/jdeq.2001.4151}}.

\bibitem{vtk-book}
W.~Schroeder, K.~Martin, B.~Lorensen, {The Visualization Toolkit, 4th edition},
  {Kitware Inc.}, 2006.

\bibitem{vtk-file-formats}
{VTK File Formats}, \url{www.vtk.org/VTK/img/file-formats.pdf} (2015).

\bibitem{Bernal2013}
L.M. Bernal, V.M. Calo, N.~Collier, G.A. Espinosa, F.~Fuentes, J.C. Mahecha,
  Isogeometric analysis of hyperelastic materials using {PetIGA}, Procedia
  Computer Science 18~(0) (2013) 1604--1613, 2013 International Conference on
  Computational Science.
\newblock \href {http://dx.doi.org/10.1016/j.procs.2013.05.328}
  {\path{doi:10.1016/j.procs.2013.05.328}}.

\bibitem{SimoHughes}
J.C. Simo, T.J.R. Hughes, Computational Inelasticity, Springer, 1998.

\bibitem{ContMech}
O.~Gonzalez, A.M. Stuart, A First Course in Continuum Mechanics, Cambridge
  University Press, 2008.

\bibitem{Jansen2000}
K.E. Jansen, C.H. Whiting, G.M. Hulbert, A generalized-$\alpha$ method for
  integrating the filtered {Navier--Stokes} equations with a stabilized finite
  element method, Computer Methods in Applied Mechanics and Engineering
  190~(3--4) (2000) 305--319.
\newblock \href {http://dx.doi.org/10.1016/S0045-7825(00)00203-6}
  {\path{doi:10.1016/S0045-7825(00)00203-6}}.

\bibitem{Bazilevs2007}
Y.~Bazilevs, V.M. Calo, J.A. Cottrell, T.J.R. Hughes, A.~Reali, G.~Scovazzi,
  Variational multiscale residual-based turbulence modeling for large eddy
  simulation of incompressible flows, Computer Methods in Applied Mechanics and
  Engineering 197~(1--4) (2007) 173--201.
\newblock \href {http://dx.doi.org/10.1016/j.cma.2007.07.016}
  {\path{doi:10.1016/j.cma.2007.07.016}}.

\bibitem{Motlagh2013}
Y.G. Motlagh, H.T. Ahn, T.J.R. Hughes, V.M. Calo, Simulation of laminar and
  turbulent concentric pipe flows with the isogeometric variational multiscale
  method, Computers \& Fluids 71 (2013) 146--155.
\newblock \href {http://dx.doi.org/10.1016/j.compfluid.2012.09.006}
  {\path{doi:10.1016/j.compfluid.2012.09.006}}.

\bibitem{tacc}
\href{http://www.tacc.utexas.edu/}{{Texas Advanced Computing Center}} [online]
  (2015).
\newline\urlprefix\url{http://www.tacc.utexas.edu/}

\bibitem{Wozniak2014}
M.~Wo\'{z}niak, K.~Ku\'{z}nik, M.~Paszy\'{n}ski, V.M. Calo, D.~Pardo,
  Computational cost estimates for parallel shared memory isogeometric
  multi-frontal solvers, Computers \& Mathematics with Applications 67~(10)
  (2014) 1864--1883.
\newblock \href {http://dx.doi.org/10.1016/j.camwa.2014.03.017}
  {\path{doi:10.1016/j.camwa.2014.03.017}}.

\bibitem{Yokota2013}
R.~Yokota, J.~Pestana, H.~Ibeid, D.~Keyes, Fast multipole preconditioners for
  sparse matrices arising from elliptic equations, Computing Research
  Repository (2014) 1--32.\href {http://arxiv.org/abs/1308.3339}
  {\path{arXiv:1308.3339}}.

\bibitem{Collier2015}
N.~Collier, A-.L. Haji-Ali, F.~Nobile, E.~von Schwerin, R.~Tempone, A
  continuation multilevel {M}onte {C}arlo algorithm, BIT Numerical Mathematics
  55~(2) (2015) 399--432.
\newblock \href {http://dx.doi.org/10.1007/s10543-014-0511-3}
  {\path{doi:10.1007/s10543-014-0511-3}}.

\bibitem{Casquero2015}
H.~Casquero, C.~Bona-Casas, H.~Gomez, A {NURBS}-based immersed methodology for
  fluid-structure interaction, Computer Methods in Applied Mechanics and
  Engineering 284~(0) (2015) 943--970, isogeometric Analysis Special Issue.
\newblock \href {http://dx.doi.org/10.1016/j.cma.2014.10.055}
  {\path{doi:10.1016/j.cma.2014.10.055}}.

\bibitem{Rudraraju2014}
S.~Rudraraju, A.~Van der Ven, K.~Garikipati, Three-dimensional isogeometric
  solutions to general boundary value problems of toupin’s gradient
  elasticity theory at finite strains, Computer Methods in Applied Mechanics
  and Engineering 278~(0) (2014) 705--728.
\newblock \href {http://dx.doi.org/10.1016/j.cma.2014.06.015}
  {\path{doi:10.1016/j.cma.2014.06.015}}.

\bibitem{Vignal2015}
P.~Vignal, A.~Sarmiento, A.M.A. C\^{o}rtes, L.~Dalcin, V.M. Calo, Coupling
  {N}avier-{S}tokes and {C}ahn-{H}illiard equations in a two-dimensional
  annular flow configuration, Procedia Computer Science 51~(0) (2015) 934--943,
  international Conference On Computational Science, \{ICCS\} 2015
  Computational Science at the Gates of Nature.
\newblock \href {http://dx.doi.org/10.1016/j.procs.2015.05.228}
  {\path{doi:10.1016/j.procs.2015.05.228}}.

\bibitem{numpy}
T.E. Oliphant, A Guide to NumPy, Trelgol Publishing, 2006.

\bibitem{matplotlib}
J.D. Hunter, Matplotlib: A {2D} graphics environment, Computing in Science \&
  Engineering 9~(3) (2007) 90---95.
\newblock \href {http://dx.doi.org/10.1109/MCSE.2007.55}
  {\path{doi:10.1109/MCSE.2007.55}}.

\bibitem{ipython}
F.~P\'erez, B.E. Granger, {IP}ython: a system for interactive scientific
  computing, Computing in Science \& Engineering 9~(3) (2007) 21--29.
\newblock \href {http://dx.doi.org/10.1109/MCSE.2007.53}
  {\path{doi:10.1109/MCSE.2007.53}}.

\bibitem{paraview}
A.~Henderson, {ParaView} guide, a parallel visualization application, Tech.
  Rep. Revision 4.1, Kitware Inc. (2014).

\end{thebibliography}

\end{document}